\documentclass[onecolumn,nofootinbib,preprintnumbers, sort&compress, floatfix,a4paper,superscriptaddress,9pt]{revtex4-1}
\usepackage[framemethod=TikZ]{mdframed}
\usepackage[most]{tcolorbox}
\usepackage{longtable}
\usepackage{mathcomp}
\usepackage{pifont}
\usepackage{xcolor}
\usepackage{amsthm}
\usepackage{pgf}
\usepackage{pgfpages}
\usepackage{amsmath}
\usepackage{amsfonts}
\usepackage{amssymb}
\usepackage{graphicx}
\usepackage{bm}
\usepackage{hyperref}
\usepackage{lipsum}
\usepackage{array}
\usepackage{booktabs}
\usepackage{tikz}
\usepackage{titlesec}
\usepackage{hyperref}
\hypersetup{bookmarksnumbered}
\def\be{\begin{equation}}
\def\ee{\end{equation}}
\def\bi{\bibitem}
\usepackage{amsmath}
\usepackage{empheq}
\usepackage[most]{tcolorbox}
\newtcbox{\mymath}[1][]{%
    nobeforeafter, math upper, tcbox raise base,
    enhanced, colframe=blue!30!black,
    colback=blue!30, boxrule=1pt,
    #1}
\usepackage{mdframed}
\usepackage{lipsum}
\usepackage{PGF}
\usepackage{pgfkeys}
\usepackage{pgffor}
\usepackage{pgfcalendar}
\usepackage{pgfpages}
\usepackage{tikz}
\usetikzlibrary{calc}
\usepackage{array}
\usepackage{booktabs}
\usepackage{multirow}
\usepackage{latexsym}
\usepackage[english]{babel}
\allowdisplaybreaks
\vbox{\hbox{1000}}
\hypersetup{colorlinks=true,linkcolor=blue!100,citecolor=red!100}
\newcounter{theo}[section] 
\renewcommand{\thetheo}{\arabic{section}.\arabic{theo}}

\newcounter{prf}[section]
\renewcommand{\theprf}{\arabic{section}.\arabic{prf}}

\newcounter{lem}[section]
\renewcommand{\thelem}{\arabic{section}.\arabic{lem}}

\titleformat{\section}
{\color{black}\normalfont\Large\bfseries}
{\color{black}\thesection}{1em}{}
\renewcommand\thesection{\arabic{section}}

\newcommand*{\hrmybox}[2]{\colorbox[rgb]{0.70,0.70,1.00}{\parbox{.99\linewidth}{#2}}}
\newcommand*{\rmybox}[2]{\colorbox[rgb]{0.91,0.45,0.45}{\parbox{.99\linewidth}{#2}}}

\newcommand*{\pqrmybox}[2]{\colorbox[rgb]{0.00,0.98,0.25}{\parbox{.99\linewidth}{#2}}}
\newcommand*{\goldmybox}[2]{\colorbox[rgb]{0.76,0.98,0.25}{\parbox{.99\linewidth}{#2}}}

\allowdisplaybreaks
\usepackage{booktabs}

\begin{document}
\title{\textcolor[rgb]{0.00,0.40,0.80}{\textsf{\huge\boldmath
Late-time-accelerated expansion arisen from gauge fields in an anisotropic background and a fruitful trick for Noether's approach}}}
\author{‪Behzad Tajahmad}
\email{behzadtajahmad@yahoo.com}
\affiliation{Faculty of Physics, University of Tabriz, Tabriz, Iran}
\affiliation{Research Institute for Astronomy and Astrophysics of Maragha (RIAAM)-Maragha, Iran, P.O. Box: 55134-441\\}
\begin{center}
\begin{abstract}
\begin{tcolorbox}[breakable,colback=white,
colframe=cyan,width=\dimexpr\textwidth+0mm\relax,enlarge left by=-17mm,enlarge right by=-6mm ]
\text{ }\\
\large{\textbf{\textsf{Abstract:}}
In this paper, a modified teleparallel gravity action containing a coupling between a scalar field potential and magnetism, in anisotropic and homogeneous backgrounds, is investigated through Noether symmetry approach. The focus of this work is to describe late-time-accelerated expansion.\\
Since finding analytical solutions carrying all conserved currents emerged by Noether symmetry approach, is very difficult, hence regularly in the literature, the authors split the total symmetry into sub-symmetries and then select, usually, some of them to be carried by the solutions. This manner limits the forms of unknown functions obtained. However, in ref. \cite{g51}, B.N.S. approach was proposed in order to solve such problems but its main motivation was carrying more conserved currents by solutions. In this paper, by eliminating the aforementioned limitation as much as possible, a trick leading to some graceful forms of unknown functions is suggested. Through this fruitful approach, the solutions may carry more conserved currents than usual ways and maybe new forms of symmetries. I named this new approach to be CSSS-trick (\textbf{C}ombination of \textbf{S}ub-symmetries through \textbf{S}pecial \textbf{S}elections). With this approach, it is demonstrated that the unified dark matter potential is deduced by the gauge fields.\\
Utilizing the $\mathfrak{B}\text{-function}$ method, a detailed data analysis of results obtained yielding perfect agreements with recent observational data are performed.\\
And finally, the Wheeler-De Witt (WDW) equation is discussed to demonstrate recovering the Hartle criterion due to the oscillating feature of the wave function of the universe.}
\end{tcolorbox}
\end{abstract}

\end{center}
\maketitle
\hrule \hrule \hrule
\textbf{\textcolor[rgb]{0.00,0.00,0.00}{\tableofcontents}}
\text{ }
\hrule \hrule \hrule
\noindent\hrmybox{}{\section{Introduction\label{sec:intro}}}\vspace{5mm}

One of the major challenges for physicists is the explanation of the essence and mechanism of the acceleration of our universe. Accelerated expansion of the universe has been confirmed by several astrophysical observations including supernova type Ia \cite{sup1,sup2}, weak lensing \cite{lens}, CMB studies \cite{CMB}, baryon acoustic oscillations \cite{baryon}, and large-scale structure \cite{LSC}. This discovery is inconsistent with the standard Einstein's general relativity. In general, two main classes of ideas (solutions) to understand the late-time-accelerated expansion have been proposed. Since this acceleration needs to negative pressure to occur, hence the first approach is the existence of an exotic liquid, so-called dark energy that about $70\%$ of the universe is made up of it. The most probable solution to dark energy was thought that is Einstein's cosmological constant \cite{Sol-DE}, but it failed because it cannot resolve `fine tuning' and `cosmic coincidence' problems. Hence, other theoretical models such as the phantom field \cite{phantom1, phantom2, phantom3, phantom4, phantom5, phantom6}, quintom \cite{quintom1, quintom2, quintom3, quintom4}, quintessence \cite{quin1, quin2, quin3}, and tachyon field \cite{tachyon} have been suggested. The second possibility is to modify Einstein's general relativity \cite{MG1, MG2} making the action of the theory dependent upon a function of the curvature scalar $R$. In a certain limit of the parameters, as we expect, the theory reduces to general relativity. Recently, various novel gravitational modification theories like scalar-tensor theories, $f(R)$-gravity, $f(T)$-gravity, $f(T)$-gravity with boundary term ($f(T,B)$) \cite{bahabah}, $f(T)$-gravity with an unusual term \cite{beh-intervention} and etcetera have been suggested.

In the actions of extended theories of gravity, there are unknown functions. The choice of the unknown functions, somewhat arbitrary, has given rise to the objection of fine-tuning, the very problem whose solutions have been set out through inflationary theories. Therefore, it is desirable to have a standard path to extract unknown functions (especially the potential) of extended theories of gravity. One such approach is based upon the celebrated Noether symmetry approach and it was applied by many authors (for example, see refs. \cite{g51,no2,no3,no3.1,no3.2,no4,no5,no6,no7,no8,no9,no10,no11,no12,no13,
no14,no15,no16, no17,Y1no17,Y2no17,Y3no17,no18,no19,no20,no21,no22,no23,no24,no25,no26,no27,
no28,no29,no30,no31, no32,no33,no34,no35,no36,no37,no38}). Noether symmetry approach
enables one to find out conserved quantities from the presence of variational symmetries \cite{g16}. However, some hidden conserved currents may not be obtained by the Noether symmetry approach \cite{g17, g31}. Furthermore, this approach may fail to get the purpose (Finding solutions whose carry all conserved currents or at least more of those which obtained by Noether symmetry approach), hence the B.N.S. approach has recently been suggested \cite{g51}.

Before terminating this section, let us present a short review of the Noether symmetry approach from Prof. S. Capozziello's papers on this subject (For more and complete information see, for example, refs. \cite{00z10,00z11}).

Let $\mathcal{L}(q^{i},\dot{q}^{i})$ be a canonical, non degenerate point-like Lagrangian satisfying
\begin{align}\label{no1}
\frac{\partial \mathcal{L}}{\partial \lambda}=0, \qquad \det \left(H_{ij} \right)\equiv \det \left\|\frac{\partial^2 \mathcal{L}}{\partial \dot{q}^{i} \partial \dot{q}^{j}} \right\| \neq0,
\end{align}
where $H_{ij}$ is the Hessian matrix of the Lagrangian and a dot indicates derivative with respect to the affine parameter $\lambda$ which usually corresponds to the cosmic time $t$. The Lagrangian in analytic mechanics is of the form
\begin{align}\label{no2}
\mathcal{L}=E_{k.}(\mathbf{q},\dot{\mathbf{q}})-V(\mathbf{q}),
\end{align}
where $E_{k}$ and $V$ are the positive definite quadratic kinetic energy and potential energy, respectively. The Hamiltonian associated with $\mathcal{L}$ is:
\begin{align}\label{no3}
E_{\mathcal{L}}\equiv \frac{\partial \mathcal{L}}{\partial \dot{q}^{i}}-\mathcal{L},
\end{align}
it coincides with the total energy $E_{k}+V$, and is a constant of motion. Note that this constant of motion is a fruit of (complete) Noether symmetry approach when the point-like Lagrangian does not explicitly depend upon time and its generator is $\partial / \partial t$. The prevalent cosmological problems have a finite number degrees of freedom, hence the point transformations are considered. Any smooth and invertible transformation of the generalized coordinates $q^{i} \to Q^{i}(\mathbf{q})$ induces a transformation of the generalized velocities
\begin{align}\label{no4}
\dot{q}^{i} \rightarrow \dot{Q}^{i}(\mathbf{q})=\frac{\partial Q^{i}}{\partial q^{j}} \dot{q}^{j}.
\end{align}
the matrix $\mathcal{J}=\|\partial Q^{i}/ \partial q^{j} \|$ is the Jacobian of the transformation on the positions, and it is assumed to be nonzero. A point transformation $Q^{i}=Q^{i}(\mathbf{q})$ may depend upon one or more parameters. One may suppose that a point transformation depends upon a parameter, therefore, the transformation is then generated by a vector field.
In general, an infinitesimal point transformation is represented by a generic vector field on $Q$
\begin{align}\label{no5}
\mathbf{X}=\alpha^{i}(\mathbf{q})\frac{\partial}{\partial q^{i}}.
\end{align}
The induced transformation (\ref{no4}) is then represented by
\begin{align}\label{no6}
\mathbf{X}^{(c)}=\alpha^{i}(\mathbf{q}) \frac{\partial}{\partial q^{i}}+\left(\frac{\mathrm{d}}{\mathrm{d}\lambda}\alpha^{i}(\mathbf{q}) \right)\frac{\partial}{\partial \dot{q}^{i}}.
\end{align}
A function $F(\mathbf{q},\dot{\mathbf{q}})$ is invariant under the transformation $\mathbf{X}^{(c)}$ if
\begin{align}\label{no7}
\mathfrak{L}_{\mathbf{X}^{(c)}}F \equiv \alpha^{i}(\mathbf{q})\frac{\partial F}{\partial q^{i}}+\left(\frac{\mathrm{d}}{\mathrm{d}\lambda}\alpha^{i}(\mathbf{q}) \right)\frac{\partial F}{\partial \dot{q}^{j}}=0,
\end{align}
where $\mathfrak{L}_{\mathbf{X}^{(c)}}$ is the Lie derivative of $F$ along $\mathbf{X}^{(c)}$.  If $\mathfrak{L}_{\mathbf{X}^{(c)}}=0$, $\mathbf{X}^{(c)}$ is then a symmetry for the dynamics derived by $\mathcal{L}$. Now, we consider a Lagrangian $\mathcal{L}$ leading to the Euler-Lagrange equations
\begin{align}\label{no8}
\frac{\mathrm{d}}{\mathrm{d}\lambda}\frac{\partial \mathcal{L}}{\partial \dot{q}^{j}}-\frac{\partial \mathcal{L}}{\partial q^{j}}=0.
\end{align}
and the vector field (\ref{no6}) which is called the complete lift of $\mathbf{X}$. Contracting (\ref{no8}) with $\alpha^{i}\text{'s}$ yields
\begin{align}\label{no9}
\alpha^{j} \left[\frac{\mathrm{d}}{\mathrm{d}\lambda}\frac{\partial \mathcal{L}}{\partial \dot{q}^{j}}-\frac{\partial \mathcal{L}}{\partial q^{j}} \right]=0.
\end{align}
Using the total derivative relation as
\begin{align}\label{no10}
\alpha^{j}\frac{\mathrm{d}}{\mathrm{d}\lambda}\left(\frac{\partial \mathcal{L}}{\partial \dot{q}^{j}} \right)=\frac{\mathrm{d}}{\mathrm{d}\lambda}\left(\alpha^{j}\frac{\partial \mathcal{L}}{\partial \dot{q}^{j}} \right)-\left(\frac{\mathrm{d}\alpha^{j}}{\mathrm{d}\lambda} \right)\frac{\partial \mathcal{L}}{\partial \dot{q}^{j}},
\end{align}
one obtains from equation (\ref{no9}) that
\begin{align}\label{no11}
\frac{\mathrm{d}}{\mathrm{d}\lambda}\left(\alpha^{i}\frac{\partial \mathcal{L}}{\partial \dot{q}^{i}} \right)=\mathfrak{L}_{\mathbf{X^{(c)}}}\mathcal{L}.
\end{align}
Noether theorem is a straightforward consequence of this equation. The Noether theorem states that if $\mathfrak{L}_{\mathbf{X^{(c)}}}\mathcal{L}=0$, then the function
\begin{align}\label{no12}
\Sigma_{0}=\alpha^{k}\frac{\partial \mathcal{L}}{\partial \dot{q}^{k}}
\end{align}
is a constant of motion. It is worth noting that eq. (\ref{no12}) may be expressed in a coordinate-independent way as the contraction of $\mathbf{X}$ with the Cartan one-form
\begin{align}\label{no13}
\theta_{\mathcal{L}}\equiv \frac{\partial \mathcal{L}}{\partial \dot{q}^{i}}\mathrm{d}q^{i}.
\end{align}
Given a generic vector field $\mathbf{Y}=y^{i}\partial / \partial x^{i}$ and 1-form $\beta = \beta_{i}\mathrm{d}x^{i}$ it is, by definition, $i_{\mathbf{Y}}\beta=y^{i}\beta_{i}$, and eq. (\ref{no12}) can then be expressed as
\begin{align}\label{no14}
i_{\mathbf{X}}\theta_{\mathcal{L}}=\Sigma_{0}
\end{align}
By a point-transformation, the vector field $\mathbf{X}^{(c)}$ becomes
\begin{align}\label{no15}
\tilde{\mathbf{X}}^{(c)}=\left(i_{\mathbf{X}}\mathrm{d}Q^{k} \right)\frac{\partial}{\partial Q^{k}}+\left[\frac{\mathrm{d}}{\mathrm{d}\lambda}
\left(i_{\mathbf{X}}\mathrm{d}Q^{k} \right) \right]\frac{\partial}{\partial \dot{Q}^{k}}.
\end{align}
$\tilde{\mathbf{X}}^{(c)}$ is still the lift of a vector field defined on the space of positions (configuration space). If $\mathbf{X}$ is a symmetry and we choose a point transformation such that
\begin{align}\label{no16}
i_{\mathbf{X}}\mathrm{d}Q^{1}=1; \qquad i_{\mathbf{X}}\mathrm{d}Q^{i}=0\quad i\neq 1,
\end{align}
we get
\begin{align}\label{no17}
\tilde{\mathbf{X}}^{(c)}=\frac{\partial}{\partial Q^{1}}; \qquad \frac{\partial \mathcal{L}}{\partial Q^{1}}=0.
\end{align}
Therefore, $Q^{1}$ is a cyclic coordinate and the dynamics can be reduced. The coordinate transformation (\ref{no16}) is not unique and a clever choice is very important part of this procedure as it can be so advantageous. In general, the solution of equation (\ref{no16}) is not defined on the whole space, rather, it is local. The important point which is also used in this paper is that in the case of multiple vector fields $\mathbf{X}$, say $\mathbf{X}_{1}$ and $\mathbf{X}_{2}$, if these commute, i.e. $\left[ \mathbf{X}_{1},\mathbf{X}_{2} \right]=0$, then two cyclic coordinates can be found by solving the following system
\begin{align}\label{cappp}
i_{\mathbf{X}_{1}}\mathrm{d}Q^{1}=1, \quad
i_{\mathbf{X}_{2}}\mathrm{d}Q^{2}=1, \quad
i_{\mathbf{X}_{1}}\mathrm{d}Q^{i}=0\quad (i\neq 1), \quad
i_{\mathbf{X}_{2}}\mathrm{d}Q^{i}=0\quad (i\neq 2).
\end{align}
Hence, $\partial / \partial Q^{1}$ and $\partial / \partial Q^{2}$ would be the transformed fields. Because in the current problem of study, our symmetry generators commute with each other, hence we do not review what we should do if they do not commute.\\

\noindent\hrmybox{}{\section{The model and field equations \label{II}}}\vspace{5mm}

We start with the gravitational action of the form \cite{g51}
\begin{align}\label{action}
S=\int d^{4}x \; e \left[\frac{M^{2}_{\mathrm{Pl}}}{2} \; T +\frac{1}{2}\varphi_{,\mu} \varphi^{,\mu}-V(\varphi)-\frac{1}{4}f^{2}(\varphi) F_{\mu \nu} F^{\mu \nu} \right],
\end{align}
where $e=\mathrm{det}(e^{i}_{\nu})=\sqrt{-g}$ with $e^{i}_{\nu}$ being a vierbein (tetrad) basis, $M_{\mathrm{Pl}}$ is the reduced Planck mass, $T$ is the torsion scalar, $\varphi_{\mu}$ stands for the components of the gradient of $\varphi(t)$, $V(\varphi)$ is the scalar field potential, and $f^{2}(\varphi)$ is the gauge kinetic function that has been coupled to the strength tensor $F_{\mu \nu}$. The electromagnetic field tensor $\mathbf{F}$ is generated by the vector potential $\mathbf{A}$ of electromagnetic theory through the geometric relation $\mathbf{F}= - \text{(antisymmetric part of $\nabla \mathbf{A}$)}$. Hence, for a given 4-potential $A_{\mu}$, the field strength of the vector field is defined by $F_{\mu \nu}=\partial_{\nu} A_{\mu}-\partial_{\nu} A_{\mu} \equiv A_{\nu,\mu} - A_{\mu,\nu}$.\\
In several papers such as refs. \cite{NN1,NN2,NN3,NN5,g51}, the action (\ref{action}) was investigated\footnote{Note that both actions $\int d^{4}x \; \sqrt{-g}[R+\cdots]$ and $\int d^{4}x \; e [T+ \cdots]$ lead to the same field equations.}, the studies of which led to satisfactory results especially describing early inflation and late-time-accelerated expansion. The action (\ref{action}) is the most generic action for single field inflation. However, there is room to make some `trivial' generalizations like adding a scalar field coupling function to torsion and etcetera, but the basic is the action of the form (\ref{action}). The success of the action (\ref{action}) in the elucidation of the inflation era is due to the fact that the gauge fields are the main driving force for the inflationary background. It is worth mentioning that there are several fields, such as the vector fields and the nonlinear electromagnetic fields, which are able to produce negative pressure effects. On the other hand, the accelerating picture of the expanding nature of the universe requires a negative pressure. Hence, it is a substantial motivation to obtain a unified model (with a single scalar field) by action (\ref{action}) which describes the stages of cosmic evolution.\\
In this paper, we want to answer the question of whether or not this model may describe the late-time-accelerated expansion in the anisotropic and homogeneous background, namely Locally Rotationally Symmetric Bianchi type I (LRS B-I).\\
The LRS B-I line element is given by
\begin{align}\label{line element}
\mathrm{d}s^2=\mathrm{d}t^2-a^2(t)\mathrm{d}x^2-b^2(t) \left[\mathrm{d}y^2+\mathrm{d}z^2\right],
\end{align}
where the expansion radii $a$ and $b$ are functions of time $t$. Therefore, the torsion scalar for this background turns out to be
\begin{align}\label{torsion}
T= -2 \left(2\frac{\dot{a}}{a}\frac{\dot{b}}{b} +\frac{\dot{b}^2}{b^2} \right)
=-2 \left(2H_{1}H_{2}+H^2_{2}\right),
\end{align}
where the dot denotes a differentiation with respect to time and $H_{1}$, and $H_{2}$ are the directional Hubble parameters ($H_{1}$ along $x$ direction while $H_{2}$ along $y$ and $z$ directions).\\
In a spatially homogeneous model the ratio of shear scalar $\sigma$,
\begin{align}\label{zub-sig}
\sigma^2=\frac{1}{2}\sigma_{AB}\sigma^{AB}=\frac{1}{3}
\left(\frac{\dot{a}}{a}-\frac{\dot{b}}{b} \right)^2,
\end{align}
to expansion scalar $\Theta$,
\begin{align}\label{zub-Thet}
\Theta=u^{\lambda}_{;\lambda}=\frac{\dot{a}}{a}+2\frac{\dot{b}}{b},
\end{align}
is constant (i.e. $\sigma / \Theta=constant$). This compels the condition $a=b^{m}$ with $m\neq 0$. Manifestly, $m=0$ is nonphysical because it means that one of the scale factors is constant (i.e. $a=1$), and $m=1$ is flat FRW space-time. It has been demonstrated in ref. \cite{beh20192} that according to recent observational data, $m$ is very close to $1$. Utilizing this well-known reasonable condition, $a=b^{m}$, the expansion scalar (\ref{torsion}) takes the form
\begin{align}
T=-2(2m+1) \frac{\dot{b}^2}{b^2}=-2(2m+1)H^2_{2}. \label{torsion1}
\end{align}
Regarding (\ref{line element}), we introduce the homogeneous and anisotropic vector field as
\begin{align}\label{AMU}
A_{\mu}=\left(A_{0};A_{1},A_{2},A_{3} \right)
=\left(\chi(t);k_{1}A(t),\frac{k_{2}}{\sqrt{2}}A(t),\frac{k_{2}}{\sqrt{2}}A(t) \right),
\end{align}
whence we get
\begin{align}\label{FMU}
F_{\mu \nu}F^{\mu \nu}=-2\left(\frac{k^{2}_{1}}{b^{2m}}+\frac{k^{2}_{2}}{b^2} \right) \dot{A}^2.
\end{align}
One may choose the gauge $A_{0}=\chi(t)=0$, by using the gauge invariance \cite{NN5}. For simplicity, let us assume that the direction of the vector field does not change in time. Pursuant to the background geometry (\ref{line element}), generally $k_{1} \neq k_{2}/\sqrt{2}$. The special case $k_{1} = k_{2}/\sqrt{2}$ is for FRW space-time. Furthermore, it is readily observed that one may take one of $k_{1}$ or $k_{2}$ equal to zero. But, we prefer to keep both. However, in section (\ref{sect3}), it is indicated that the Noether symmetry approach does not allow to maintain both, hence it generates two classes.\\
Writing the action (\ref{action}) in the canonical form $S=\int \mathrm{d}t \; \mathcal{L}(Q,\dot{Q})+\Sigma_{0}$ down, the point-like Lagrangian would be\footnote{In our case, the surface-term $\Sigma_{0}$ is zero.}
\begin{align}\label{point-like(LRS)}
\mathcal{L}=(2m+1)b^m \dot{b}^2-\frac{1}{2}b^{m+2} \dot{\varphi}^2+b^{m+2}V-\frac{1}{2}k^{2}_{1}b^{2-m} f^2\dot{A}^2-\frac{1}{2}k^{2}_{2}b^{m} f^2\dot{A}^2,
\end{align}
where the reduced Planck mass $M_{\mathrm{Pl}}$ has been set equal to $1$.\\
The Euler-Lagrange equations for a dynamical system are given by
\begin{align}\label{ele1}
\frac{\partial \mathcal{L}}{\partial q_{i}}-\frac{\mathrm{d}}{\mathrm{d}t}\left(\frac{\partial \mathcal{L}}{\partial \dot{q}_{i}} \right)=0,
\end{align}
in which $q_{i}$ are the generalized positions in the corresponding configuration space $Q=\{q_{i}\}$. According to (\ref{point-like(LRS)}), our configuration space reads $Q=\{b, \varphi, A\}$ and consequently, its tangent space would be $TQ=\{b, \dot{b}, \varphi, \dot{\varphi}, A, \dot{A}\}$.\\
Pursuant to the point-like Lagrangian (\ref{point-like(LRS)}), the corresponding Euler-Lagrange equation for the scale factor $b$ reads
\begin{align}\label{scale factor b}
(2m+1)H_{2}^{2}+\frac{2(2m+1)}{(m+2)}\dot{H}_{2}=- \; \frac{1}{2}\dot{\varphi}^2 + V - k^{2}_{1} \; \frac{(2-m)}{2(m+2)}\frac{f^2 \dot{A}^2}{b^{2m}}- k^{2}_{2} \; \frac{m}{2(m+2)}\frac{f^2 \dot{A}^2}{b^2}.
\end{align}
For the scalar field $\varphi$, the Euler-Lagrange equation becomes
\begin{align}\label{scalar field}
\ddot{\varphi}+(m+2)H_{2}\dot{\varphi}+V^{\prime} - k^{2}_{1} \; \frac{f f^{\prime} \dot{A}^2}{b^{2m}}-k^{2}_{2} \; \frac{f f^{\prime} \dot{A}^2}{b^2}=0,
\end{align}
which is the Klein-Gordon equation. The prime indicates the derivative with respect to $\varphi$. For the vector potential $A$, the Euler-Lagrange equation takes the following form:
\begin{align}\label{4v-equation}
(k^{2}_{1}b^{2-m}+k^{2}_{2}b^{m})f^2 \ddot{A}+(k^{2}_{1}(2-m)\dot{b}b^{1-m}+k^{2}_{2}m\dot{b}b^{m-1}) f^2 \dot{A}+2(k^{2}_{1}b^{2-m}+k^{2}_{2}b^{m})f f^{\prime} \dot{\varphi} \dot{A}=0.
\end{align}
The energy function associated with a Lagrangian is given by
\begin{align}\label{ele2}
E_{\mathcal{L}}=\sum_{i} \dot{q}_{i}\frac{\partial \mathcal{L}}{\partial \dot{q}_{i}}-\mathcal{L}.
\end{align}
Therefore the Hamiltonian constraint or total energy $E_{\mathcal{L}}$ corresponding to the $\binom{0}{0}$-Einstein equation becomes
\begin{align}\label{ELL}
(2m+1)H_{2}^2=\frac{1}{2}\dot{\varphi}^2+V+k^{2}_{1} \; \frac{f^2 \dot{A}^2}{2 b^{2m}}+k^{2}_{2} \; \frac{f^2 \dot{A}^2}{2 b^2}.
\end{align}
According to (\ref{scale factor b}) and (\ref{ELL}), the effective Equation of State (EoS) parameter turns out to be:
\begin{align}\label{EOS}
W_{\mathrm{eff.}}=\frac{P_{\mathrm{eff.}}}{\rho_{\mathrm{eff.}}}=\frac{\frac{1}{2}
\dot{\varphi}^2-V+k^{2}_{1}\frac{(2-m)}{(m+2)} \; \frac{f^2 \dot{A}^2}{ 2b^{2m}}+k^{2}_{2}\frac{m}{(m+2)} \; \frac{f^2 \dot{A}^2}{ 2b^2}}{\frac{1}{2}\dot{\varphi}^2+V+ k^{2}_{1} \frac{f^2 \dot{A}^2}{2b^{2m}}+k^{2}_{2}\frac{f^2 \dot{A}^2}{2b^2}}.
\end{align}
The dynamic of our system is given by these four equations
(i.e. \ref{scale factor b}, \ref{scalar field}, \ref{4v-equation}, and \ref{ELL}). The Noether approach is used in the next section to obtain exact solutions with symmetries of the extended theory of gravity (\ref{action}).\\

\noindent\hrmybox{}{\section{Nother symmetry approach and CSSS-trick\label{sect3}}}\vspace{5mm}

In this section, solving field equations (i.e. \ref{scale factor b}, \ref{scalar field}, \ref{4v-equation}, and \ref{ELL}) in order to investigate the circumstances of some important cosmological events like late-time accelerated expansion, phase crossing, and etcetera, are desired. Finding suitable forms of the unknown functions of the action (\ref{action}) to reach the aforementioned purpose are challenging, hence exploring their forms through a `standard way' seems necessary. Furthermore, it would be very beautiful if the solutions carry some conserved currents (Symmetries) as well. To this end, we utilize the Noether symmetry approach which exactly does this job.\\
Pursuant to our tangent space of the configuration space, $TQ=\{b, \dot{b}, \varphi, \dot{\varphi}, A, \dot{A}\}$, the existence of the Noether symmetry implies the existence of a vector field $\mathbf{X}$ as
\begin{equation}\label{vector}\begin{split}
\mathbf{X}=\beta \frac{\partial}{\partial b}
&+\alpha \frac{\partial}{\partial A}
+\gamma \frac{\partial}{\partial \varphi}
+\beta_{,t} \frac{\partial}{\partial \dot{b}}
+\alpha_{,t} \frac{\partial}{\partial \dot{A}}
+\gamma_{,t} \frac{\partial}{\partial \dot{\varphi}},
\end{split}\end{equation}
where
\begin{align}\label{alk1}
y=y(b,\varphi,A) \longrightarrow y_{,t}=\dot{b} \frac{\partial y}{\partial b}+\dot{\varphi} \frac{\partial y}{\partial \varphi}+ \dot{A} \frac{\partial y}{\partial A} \; ; \qquad y \in \{\alpha, \beta, \gamma \},
\end{align}
such that
\begin{align}\label{Nother condition}
\mathfrak{L}_{\mathbf{X}}\mathcal{L}=0 \nonumber \\
\longrightarrow \beta \frac{\partial \mathcal{L}}{\partial b}
&+\alpha \frac{\partial \mathcal{L}}{\partial A}
+\gamma \frac{\partial \mathcal{L}}{\partial \varphi}
\nonumber \\ &+\left(\dot{b}\frac{\partial \beta}{\partial b}+\dot{\varphi}\frac{\partial \beta}{\partial \varphi}+\dot{A}\frac{\partial \beta}{\partial A} \right)
\left(\frac{\partial \mathcal{L}}{\partial \dot{b}} \right) \nonumber\\
&+\left(\dot{b}\frac{\partial \alpha}{\partial b}+\dot{\varphi}\frac{\partial \alpha}{\partial \varphi}+\dot{A}\frac{\partial \alpha}{\partial A} \right)
\left(\frac{\partial \mathcal{L}}{\partial \dot{A}} \right)
\nonumber \\&+\left(\dot{b}\frac{\partial \gamma}{\partial b}+\dot{\varphi}\frac{\partial \gamma}{\partial \varphi}+\dot{A}\frac{\partial \gamma}{\partial A} \right)
\left(\frac{\partial \mathcal{L}}{\partial \dot{\varphi}} \right)=0.
\end{align}
This equation yields the following system of linear partial differential equations:
\begin{align}
&\left(k^{2}_{1} \left(1-\frac{1}{2}m \right)b^{2-2m}+\frac{1}{2}m k^{2}_{2} \right)\beta f+ \left(k^{2}_{1} b^{3-2m}+k^{2}_{2}b \right)\left(\gamma f^{\prime}+f\left(\frac{\partial \alpha}{\partial A} \right) \right)=0, \label{NS1}\\
&(2m+1)m \beta +2(2m+1)b \left(\frac{\partial \beta}{\partial b} \right)=0, \label{NS2}\\
&\left(\frac{1}{2}m+1 \right) \beta + b \left(\frac{\partial \gamma}{\partial \varphi} \right)=0, \label{NS3}\\
&\left(m+2 \right) \beta V+ \gamma b V^{\prime}=0, \label{NS4}\\
&2(2m+1) \left(\frac{\partial \beta}{\partial \varphi} \right)-b^{2} \left(\frac{\partial \gamma}{\partial b} \right)=0, \label{NS5}\\
&2(2m+1) \left(\frac{\partial \beta}{\partial A} \right)- \left(k^{2}_{1} b^{2-2m}+k^{2}_{2} \right) f^{2} \left( \frac{\partial \alpha}{\partial b}\right)=0, \label{NS6}\\
&b^{2} \left(\frac{\partial \gamma}{\partial A} \right)+\left(k^{2}_{1} b^{2-2m}+k^{2}_{2} \right) f^{2} \left( \frac{\partial \alpha}{\partial \varphi}\right)=0. \label{NS7}
\end{align}
The 4-dimensional configuration space $Q=\{a,b,\varphi,A \}$ was reduced to the 3-dimensional one $Q=\{b,\varphi,A \}$ due to the physical assumption $a=b^{m}$, hence we have seven partial differential equations instead of eleven numbers.\\
Solving this system of linear partial differential equations, one may obtain
\begin{align}
&\beta=\left(c_{1} e^{\mu \varphi}+c_{2} e^{-\mu \varphi}\right)b^{\frac{-m}{2}}, \qquad \alpha=c_{4}, \label{sol for NS1}\\
&\gamma= - \sqrt{4m+2} \left(c_{1} e^{\mu \varphi}-c_{2}e^{-\mu \varphi} \right) b^{-\left(\frac{m+2}{2} \right)}, \label{sol for NS2}\\
&V(\varphi)= c_{3} \left(c_{1} e^{\mu \varphi}-c_{2} e^{- \mu \varphi} \right)^2, \label{sol for NS3} \\
&f(\varphi)= c_{5} \left(c_{1}e^{\mu \varphi}-c_{2} e^{-\mu \varphi} \right)^{n}, \label{sol for NS4}
\end{align}
where
\begin{equation}\label{mu}\begin{split}
\mu = \frac{(m+2)}{2 \sqrt{4m+2}},
\end{split}\end{equation}
and
\begin{align}\label{classn}
n=\left \{
  \begin{array}{ll}
    n_{1}=\frac{2-m}{2+m}, \quad \text{when } \; k_{1} \neq 0 \; \& \; k_{2}=0 &; \\ \\
    n_{2}=\frac{m}{2+m}, \quad \text{when } \; k_{1}=0 \; \& \; k_{2}\neq 0, &.
  \end{array}
\right.
\end{align}
As is observed, in the special case $m=1$ (FRW), both are equal: $n_{1}=n_{2}=1/3$. It is important to mention that if one wants to examine FRW-case, then he must take $k_{1}=k_{2}/\sqrt{2}$ in (\ref{AMU}). In this stage, we encounter a bifurcation in equations due to $n$, and therefore two classes are separated by it. Indeed, the suitable solutions obtained by Noether symmetry approach, do not allow to have a four-potential of the form (\ref{AMU}), hence, it is readily observed that (\ref{AMU}) must be split into two independent cases:
\begin{align}
&A_{\mu}=\left(A_{0};A_{1},0,0 \right)
=\left(\chi(t);k_{1}A(t),0,0 \right), \label{AMU01}\\
&A_{\mu}=\left(A_{0};0,A_{2},A_{3} \right)
=\left(\chi(t);0,\frac{k_{2}}{\sqrt{2}}A(t),\frac{k_{2}}{\sqrt{2}}A(t) \right). \label{AMU02}
\end{align}
Therefore, according to (\ref{sol for NS1}) and (\ref{sol for NS2}), the symmetry generator, (\ref{vector}), turns out to be
\begin{align}\label{vector2}
\mathbf{X}=\; &c_{1} \left(b^{-m/2} \; e^{\mu \varphi} \; \frac{\partial}{\partial b}-\sqrt{4m+2} \; e^{\mu \varphi} \; b^{-(m+2)/2} \; \frac{\partial}{\partial \varphi} \right)\nonumber \\+&c_{2} \left(b^{-m/2} \; e^{-\mu \varphi} \; \frac{\partial}{\partial b}+\sqrt{4m+2} \; e^{-\mu \varphi} \; b^{-(m+2)/2} \; \frac{\partial}{\partial \varphi} \right)\nonumber \\+&c_{4} \left(\frac{\partial}{\partial A} \right)\nonumber \\
+&c_{1} \left(\left(b^{-m/2} \; e^{\mu \varphi}\right)_{,t} \; \frac{\partial}{\partial \dot{b}}-\sqrt{4m+2} \; \left(e^{\mu \varphi} \; b^{-(m+2)/2}\right)_{,t} \; \frac{\partial}{\partial \dot{\varphi}} \right)\nonumber \\+&c_{2} \left(\left(b^{-m/2} \; e^{-\mu \varphi}\right)_{,t} \; \frac{\partial}{\partial \dot{b}}+\sqrt{4m+2} \; \left(e^{-\mu \varphi} \; b^{-(m+2)/2}\right)_{,t} \; \frac{\partial}{\partial \dot{\varphi}} \right).
\end{align}
For convenience, let us, from now on, write the symmetry generators on the configuration space $Q=\{b, \varphi, A\}$, not on $TQ$. Hence, (\ref{vector2}) splits into three independent generators:
\begin{align}
&\mathbf{X}_{1}= e^{\mu \varphi}\; \; b^{-m/2}\left(\frac{\partial}{\partial b} - \frac{\sqrt{4m+2}}{b} \frac{\partial}{\partial \varphi} \right), \label{X1} \\
&\mathbf{X}_{2}= e^{-\mu \varphi} \; b^{-m/2}\left(\frac{\partial}{\partial b} +\frac{\sqrt{4m+2}}{b} \frac{\partial}{\partial \varphi} \right),\label{X2} \\
&\mathbf{X}_{3}= \frac{\partial}{\partial A}.\label{X3}
\end{align}
because (\ref{vector2}) may be taken as:
\begin{align}\label{vector3}
\mathbf{X} \equiv c_{1}\mathbf{X}_{1}+c_{2}\mathbf{X}_{2}+c_{4}\mathbf{X}_{3}.
\end{align}
Consequently, the corresponding conserved currents are found to be
\begin{align}
\mathbf{I}_{1}&=b^{m/2} \; e^{\mu \varphi}\left[2(2m+1)\dot{b}+\sqrt{4m+2} \; b \dot{\varphi} \right],\label{CC1}\\
\mathbf{I}_{2}&=b^{m/2} \; e^{- \mu \varphi}\left[2(2m+1)\dot{b}-\sqrt{4m+2} \; b \dot{\varphi} \right],\label{CC2}\\
\mathbf{I}_{3}&=k_{j}b^{(2+m)n_{j}} f^{2} \dot{A},\label{CC3}
\end{align}
respectively. Note that there is no Einstein summation convention over the subscript $j$ in (\ref{CC3})\footnote{However, in general, there is an Einstein summation convention over the subscript $j$ ($j$ takes $1$ and $2$) in the last conserved current, but since it is demonstrated that the Noether symmetry approach does not allow to keep both $k_{1}$ and $k_{2}$ simultaneously, hence there is no summation convention over $j$.}. For underlining this point, let us use $\mathbf{I}_{3j}$ instead of $\mathbf{I}_{3}$.

It may easily be indicated that all the symmetries commute with each other, therefore the Lie algebra
\begin{align}\label{liebracket}
\left[\mathbf{X}_{\varsigma}, \mathbf{X}_{\tau} \right]=0, \qquad \varsigma, \tau=1,2,3,
\end{align}
is satisfied. The corresponding constants of motion also close the same algebra in terms of Poisson bracket:
\begin{align}\label{poissonbracket}
\left\{\mathbf{I}_{\varsigma}, \mathbf{I}_{\tau} \right\}=0, \qquad \varsigma, \tau=1,2,3.
\end{align}
The relation (\ref{liebracket}) is very important for us, since in what follows this point is used to obtain further suitable solutions, especially the form of interest for the potential (i.e. The unified dark matter potential).\\
The conserved current $\mathbf{I}_{3}$ is automatically carried by (\ref{4v-equation}), hence we put it aside and therefore, two conserved currents, $\mathbf{I}_{1}$ and $\mathbf{I}_{2}$, remain. Now, if we act as usual, then there are three possibilities: $1$. $\{c_{1}=0,\; c_{2} \neq 0\}$;  $2$. $\{c_{2}=0,\; c_{1} \neq 0\}$; $3$. $\{c_{1} \neq 0,\; c_{2} \neq 0\}$. The cases $1$ and $2$ are easy to be considered, but the third option is not an easy task, since its system of cyclic equations which will contain two cyclic variables cannot be solved easily. Let us do different work.\\
As we know, regularly, the forms of unknown functions of extended theories of gravity are specified by the Noether symmetry approach in which the symmetries are also obtained. But in almost all cases in the literature, we cannot obtain the solutions which carry all conserved currents or at least more of those. In order to solve this problem and also some further reasons, the B.N.S. approach was proposed (See ref. \cite{g51}). In this paper, I suggest a new approach which may be more interesting for cosmologists: \textbf{``Combination of Sub-symmetries through Special Selections''} (\textbf{CSSS}-trick). In this way, not only the solutions carry more/new conserved currents, but also this approach leads to graceful results; for example, in our case of study, the unified dark matter potentials of the forms $V=V_{0} \cosh^{2}(\mu \varphi)$ and also $V=V_{0} \sinh^{2}(\mu \varphi)$ are produced which are highly rewarding. \\

\noindent\pqrmybox{red}{\textbf{$\bullet$} \textbf{CSSS-Trick (Combination of Sub-symmetries through Special Selections):}}\\

In the Noether symmetry approach, it is usual that after finding the symmetry generator and its corresponding conserved current of the forms
\begin{align}
\mathbf{X}_{\mathrm{tot.}} &=\sum_{i}^{D} c_{i} \; \mathbf{X}_{i}, \qquad D=\text{Total number of sub-symmetries} \label{TX.1}\\
\mathbf{I}_{\mathrm{tot.}} &=\sum_{i}^{D} c_{i} \; \mathbf{I}_{i}, \label{TX.2}
\end{align}
where $c_{i}$ are constants, we split this total symmetry $\mathbf{X}_{\mathrm{tot.}}$ into sub-symmetries $\mathbf{X}_{i}$ because $\mathbf{X}_{\mathrm{tot.}}$ is a sum of $D$ independent symmetry generators and then search for the cases leading to analytical solutions which carry some of the sub-conserved currents $\mathbf{I}_{i}$. The reason for acting in such manner is that in almost all cases we cannot find analytical solutions which carry the total conserved current. I propose that instead of this work which surely leads to graceful results, we may also notice to the forms of unknown functions and then select the constant parameters in a way that they yield interesting forms for them. More precisely, first of all, after writing the total symmetry generator of the form (\ref{TX.1}), make sure that your chosen sub-symmetries, $\mathbf{X}_{i}$s, are independent from each other by considering all commutator of each two members of the set of sub-symmetries $\{\mathbf{X}_{1},\mathbf{X}_{2},\cdots,\mathbf{X}_{D}\}$. When all these commutators vanish --- i.e, the Lie algebra $\left[\mathbf{X}_{\varsigma}, \mathbf{X}_{\tau} \right]=0; \quad \varsigma, \tau \in \{1,2,\cdots,D\}$ is satisfied ---, then your selections in (\ref{TX.1}) are true. Otherwise, any nonzero commutator is also a symmetry and the procedure is repeated until the vector fields close the Lie algebra. If we did this at first, then it guarantees that no new symmetry will produce after any combination of symmetries. It will be advantageous and highly rewarding if, with tuning the constant parameters in (\ref{TX.1}) (for example $c_{1}$ and $c_{2}$ here; see (\ref{sol for NS3})--(\ref{sol for NS4}) and (\ref{vector3})), we act on a way that some interesting and well-known forms of unknown functions be achieved. Hence, we must back to the forms of unknown functions and first tune their constant parameters. Tuning the constant parameters nontrivially through special selections imply special combinations of symmetry generators and consequently the conserved currents.
This trick covers the results of the usual approach. Let me clarify this trick by an example:\\
In our case of study, we obtained:
\begin{align}
V(\varphi) &= c_{3} \left(c_{1} e^{\mu \varphi}-c_{2} e^{- \mu \varphi} \right)^2, \label{TX2.1} \\
f(\varphi) &= c_{5} \left(c_{1}e^{\mu \varphi}-c_{2} e^{-\mu \varphi} \right)^{n}, \label{TX2.2} \\
\mathbf{X}_{\mathrm{tot.}} &= c_{1} \; \mathbf{X}_{1}+c_{2} \; \mathbf{X}_{2}+c_{4} \; \mathbf{X}_{3} , \label{TX2.3} \\
\mathbf{I}_{\mathrm{tot.}} &=c_{1} \; \mathbf{I}_{1}+c_{2} \; \mathbf{I}_{2}+c_{4} \; \mathbf{I}_{3} \label{TX2.4}.
\end{align}
As already noticed, $\mathbf{I}_{3}$ is carried automatically by the field equations. However, its constant factor namely $c_{4}$ has not appeared in (\ref{TX2.1}) and (\ref{TX2.2}), hence the forms of $V$ and $f$ are not affected by it. Leave it aside. Now, if we act on the usual way, we split $\mathbf{X}$ into $\mathbf{X}_{1}$, $\mathbf{X}_{2}$, and $\mathbf{X}_{3}$. But since finding a set of a solution in which both $\mathbf{I}_{1}$ and $\mathbf{I}_{2}$ are carried, is a very hard task, hence we must put one of $c_{1}$ and $c_{2}$ equal to zero. But either we set $\{c_{1}=0,\; c_{2} \neq 0\}$ (leading to $V \approx \exp(-2 \mu \varphi)$ and $f \approx \exp(-n \mu \varphi)$) or $\{c_{1} \neq 0,\; c_{2}=0\}$ (yielding $V \approx \exp(+2 \mu \varphi)$ and $f \approx \exp(+n \mu \varphi)$), the forms of functions $V$ and $f$ are limited. The CSSS-trick suggests that instead of this work, we follow the following prescription:\\
First of all, we must notice the forms of the potential and coupling function. We should select the constant parameters appeared in the obtained forms of the potential and coupling function in a way that they lead to well-known forms for them. Then according to these selections for constant parameters, we combine the sub-symmetries and consequently sub-conserved currents. In our case, based, at least, on (\ref{TX2.1}), there are at least four well-known options:
\begin{enumerate}
	\item (Usual) Selection: $\{c_{1} \neq 0, \; c_{2}=0\}$:
	\begin{align}\label{F0pott01}
	V(\varphi)=V_{0} \exp(+2 \mu \varphi), \quad f(\varphi)=f_{0} \exp(+n \mu \varphi),
	\end{align}
	Therefore, between $\mathbf{X}_{1}$ and $\mathbf{X}_{2}$, only $\mathbf{X}_{1}$ is the symmetry of the system and consequently, $\mathbf{I}_{1}$ will be carried by the solutions.
	\item (Usual) Selection: $\{c_{1} = 0, \; c_{2} \neq 0\}$:
	\begin{align}\label{F0pott02}
	V(\varphi)=V_{0} \exp(-2 \mu \varphi), \quad f(\varphi)=f_{0} \exp(-n \mu \varphi),
	\end{align}
Therefore, between $\mathbf{X}_{1}$ and $\mathbf{X}_{2}$, only $\mathbf{X}_{2}$ is the symmetry of the system and consequently, $\mathbf{I}_{2}$ will be carried by the solutions.
	\item (Unusual) Selection: $\{c_{1} = +1/2, \; c_{2} = -1/2\}$:
	\begin{align}\label{F0pott03}
	V(\varphi)=V_{0} \cosh^{2}(\mu \varphi), \quad f(\varphi)=f_{0} \cosh^{n}(\mu \varphi),
	\end{align}
Therefore, instead of $\mathbf{X}_{1}$ and $\mathbf{X}_{2}$, the new symmetry $\mathbf{X}_{new1}=(\mathbf{X}_{1}-\mathbf{X}_{2})/2$ is the symmetry of the system and consequently, $\mathbf{I}_{new1}=(\mathbf{I}_{1}-\mathbf{I}_{2})/2$ will be carried by the solutions.
	\item (Unusual) Selection: $\{c_{1} = +1/2, \; c_{2} = +1/2\}$:
	\begin{align}\label{F0pott04}
	V(\varphi)=V_{0} \sinh^{2} (\mu \varphi), \quad f(\varphi)=f_{0} \sinh^{n} (\mu \varphi),
	\end{align}
Therefore, instead of $\mathbf{X}_{1}$ and $\mathbf{X}_{2}$, the new symmetry $\mathbf{X}_{new2}=(\mathbf{X}_{1}+\mathbf{X}_{2})/2$ is the symmetry of the system and consequently, $\mathbf{I}_{new2}=(\mathbf{I}_{1}+\mathbf{I}_{2})/2$ will be carried by the solutions.
\end{enumerate}
It must be noted that, in the CSSS-trick process, the commutators of symmetries which we want to be carried, after combination must be considered.
The remain of Noether symmetry approach namely cyclic variable process should be performed as usual with the difference that you will work with new set of symmetries. However, I think that it is better after achieving the desired forms of unknown functions, we proceed with the minimum number of symmetries, because when your system has a number of symmetries, indeed its behavior is restricted by these disciplines (symmetries) and thereby, finding analytical solution would be hard and in the most cases of interest, it is impossible. \\
In our case of study, for the third selection above we have:
\begin{align}
&\left[\mathbf{X}_{1},\mathbf{X}_{2} \right]=0, \quad \left[\mathbf{X}_{1},\mathbf{X}_{3} \right]=0, \quad \left[\mathbf{X}_{2},\mathbf{X}_{3} \right]=0, \label{TX3.1} \\
&\left[\mathbf{X}_{new1},\mathbf{X}_{3} \right]=0. \label{TX3.2}
\end{align}
The same relations hold for the fourth selection above. Indeed, (\ref{TX3.2}) is a result of (\ref{TX3.1}). Vanishing all commutators before combination guarantee vanishing the new ones after CSSS process, otherwise, it should be considered. Even though in each aforementioned selection, two symmetries exist for the system ($\mathbf{X}_{3}$ is common among them), but since $\mathbf{I}_{3}$ is carried automatically by the field equations, hence for each of four cases mentioned above, only one cyclic variable will exist. In sub-section (\ref{sub-section2}), these points are clarified. \\

\noindent\rmybox{}{\subsection{The usual cases: $\{c_{2}=0$, $c_{1} \neq 0 \}$ and $\{c_{1}=0$, $c_{2} \neq 0 \}$ \label{sub-section1}}}\vspace{5mm}
In this sub-section, two cases $\{c_{2}=0$, $c_{1} \neq 0 \}$ (\textbf{C1}) and $\{c_{1}=0$, $c_{2} \neq 0 \}$ (\textbf{C2}) are considered. It must be noted that since in each case, $\mathbf{I}_{3j}$ is carried automatically by field equations, hence we take $c_{4}=0$ in throughout this paper.\\
In order to simplify the system of equations, we use cyclic variables associated with the Noether symmetry generators $\mathbf{X}_{1}$ and $\mathbf{X}_{2}$ for cases $\{c_{2}=0$, $c_{1} \neq 0 \}$ (\textbf{C1}-class) and $\{c_{1}=0$, $c_{2} \neq 0 \}$ (\textbf{C2}-class), respectively. The existence of the Noether symmetry ensures the presence of cyclic variables, say
\begin{align}\label{ele10}
(b,\varphi, A) \longrightarrow (w,u,v),
\end{align}
where $w=w(b,\varphi, A)$, $u=u(b,\varphi, A)$, and $v=v(b,\varphi, A)$, such that the Lagrangian becomes cyclic in one of them ($w$ in our case).\\
By defining a transformation $i: (b,\varphi, A) \to (w,u,v)$ as an interior product such that
\begin{align}\label{ele11}
i_{\mathbf{X}_{1}} \mathrm{d}w=1, \quad i_{\mathbf{X}_{1}} \mathrm{d}u=0, \quad i_{\mathbf{X}_{1}} \mathrm{d}v=0
\end{align}
and
\begin{align}\label{ele12}
i_{\mathbf{X}_{2}} \mathrm{d}w=1, \quad i_{\mathbf{X}_{2}} \mathrm{d}u=0, \quad i_{\mathbf{X}_{2}} \mathrm{d}v=0
\end{align}
be held for \textbf{C1} and \textbf{C2}, respectively, the cyclic variables may then be found.\\
Solving eqs. (\ref{ele11}) and (\ref{ele12}) independently, leads to
\begin{align}\label{ac1.1}
w=\frac{b^{(m+2)/2} \; e^{-\mu \varphi}}{c_{1} (m+2)}, \quad u=\frac{b^{(m+2)/2} \; e^{\mu \varphi}}{c_{1} (m+2)}, \quad v=A,
\end{align}
and
\begin{align}\label{ac2.1}
w=\frac{b^{(m+2)/2} \; e^{\mu \varphi}}{c_{2} (m+2)}, \quad u=\frac{b^{(m+2)/2} \; e^{- \mu \varphi}}{c_{2} (m+2)}, \quad v=A,
\end{align}
for \textbf{C1} and \textbf{C2}, respectively. It is worthwhile mentioning that the choices in eqs. (\ref{ac1.1}) and (\ref{ac2.1}) are arbitrary, as more general conditions are possible. \\
Introducing a parameter $\lambda$ as
\begin{align}\label{phi123}
\left\{ \begin{array}{ll} \lambda = +1; &{} \quad \hbox {for \textbf{C1}} \; (c_{1} \neq
0, \; \hbox {and} \; c_{2}=0); \\ \\ \lambda = -1; &{} \quad \hbox {for \textbf{C2}} \; (c_{1}
=0,\; \hbox {and} \; c_{2} \neq 0),
\end{array} \right.
\end{align}
the corresponding inverse transformations would be
\begin{align}
\varphi_{1,2} &=\frac{\lambda}{2 \mu}\ln \left(\frac{u}{w} \right),  \label{phi12}\\
b_{1,2} &=\left[c_{1,2}(m+2)uw \right]^{1/(m+2)},  \label{bb123}\\
A_{1,2} &=v.  \label{bb5555123}
\end{align}
where the subscripts $1$ and $2$ correspond to \textbf{C1} and \textbf{C2} classes. As is clear from (\ref{sol for NS3}) and (\ref{sol for NS4}), the Noether symmetry approach gives the forms of the scalar field potential $V(\varphi)$ and the coupling function $f(\varphi)$ as
\begin{align}\label{pot12}
V(\varphi)=V_{0} \; \exp \left(2 \mu \lambda \varphi \right),
\end{align}
and
\begin{align}\label{cop12}
f(\varphi)=f_{0} \; \exp \left(n \mu \lambda \varphi\right),
\end{align}
where
\begin{align}
V_{0} &=c_{3} \; c_{1,2} \; \lambda,  \label{pot12.1}\\
f_{0} &=c_{5} \; c_{1,2} \; \lambda,  \label{cop12.1}
\end{align}
for both \textbf{C1} and \textbf{C2}. Therefore, according to (\ref{phi12}) and (\ref{bb123}), they are translated as follows:
\begin{align}
V(u,w) &=V_{0} \left(\frac{u}{w} \right),  \label{pot12.3}\\
f(u,w) &=f_{0} \left(\frac{u}{w} \right)^{n/2},  \label{cop12.3}
\end{align}
for both \textbf{C1} and \textbf{C2}. Therefore, the point-like Lagrangian (\ref{point-like(LRS)}) in terms of the new variables then reads
\begin{align}\label{rpointlike}
\mathcal{L}_{1,2}=&k^{2}_{1}\left(\frac{-f^{2}_{0}}{2} \left[c_{1,2} (m+2) \right]^{\frac{2(2-m)}{2+m}} \right) u^{\frac{2(2-m)}{2+m}} \dot{A}^{2}+k^{2}_{2} \left( \frac{-f^{2}_{0}}{2} \left[c_{1,2} (m+2) \right]^{\frac{2m}{2+m}}\right) u^{\frac{2m}{2+m}} \dot{A}^{2} \nonumber \\ &+\left(4c^{2}_{1,2} (2m+1) \right)\dot{u} \dot{w} +\left( V_{0}c^{2}_{1,2} (m+2)^{2}\right) u^{2}.
\end{align}
The subscripts $1$ and $2$ refer to the type of class. As already noticed, the Noether symmetry approach does not allow to keep both $k_{1}$ and $k_{2}$ nonzero simultaneously, hence the point-like Lagrangians (\ref{rpointlike}) must be decomposed into following Lagrangians:
\begin{align}\label{ele1101}
\mathcal{L}_{1j,2j}=k^{2}_{j} \frac{l_{1}}{2n_{j}} u^{2n_{j}-1} \dot{A}^{2}-l_{3}\dot{u} \dot{w}+\frac{l_{2}}{2} u^{2},
\end{align}
where the subscript $j$ can only take $j=1$ and $j=2$ corresponding to $\{k_{1} \neq 0 \; \& \; k_{2}=0\}$ and $\{k_{1}=0 \; \& \; k_{2}\neq 0 \}$, respectively, and we have defined
\begin{align}
l_{1} &=-n_{j}f^{2}_{0} \left[c_{1,2} (m+2)\right]^{2n_{j}},\\
l_{2} &=2 V_{0} c^{2}_{1,2} (m+2)^{2},\\
l_{3} &=-4c^{2}_{1,2}(2m+1),
\end{align}
in which we have used (\ref{classn}). Note that there is no Einstein's summation rule over the subscript $j$ in (\ref{ele1101}).
Both point-like Lagrangians (\ref{ele1101}) lead to the following Euler-Lagrange equations with respect to $w$, $A$, and $u$, respectively:
\begin{align}
&\ddot{u}=0,\label{rFE2}\\
&2n_{j} \dot{u} \dot{A}+u \ddot{A}=0 \label{rFE3},\\
&l_{1} k^{2}_{j} u^{2n_{j}-1} \dot{A}^{2}+l_{2}u+l_{3} \ddot{w}=0. \label{rFE1}
\end{align}
The corresponding conserved currents, (\ref{CC1}) and (\ref{CC2}), turn out to be
\begin{align}\label{reI}
\mathbf{I}_{1,2}=4(2m+1)c_{1,2} \dot{u},
\end{align}
which are equivalent to (\ref{rFE2}) (i.e. $(\mathrm{d}\mathbf{I}_{1,2}/\mathrm{d}t)=0 \equiv \ddot{u}$), hence this equation does not add any new equation.\\
Solutions to (\ref{rFE2})--(\ref{rFE1}) are
\begin{align}
u(t)&=c_{5}t+c_{6},  \label{sol1}\\
A(t)&=c_{3}+c_{4} \int u^{-2n_{j}} \mathrm{d}t,  \label{sol2}\\
w(t)&=c_{1}t+c_{2}+\int \left[ \int \left[\left(\frac{l_{1}}{l_{3}} \right) u^{2n_{j}-1} \dot{A}^2 +\left(\frac{l_{2}}{l_{3}} \right) u \right] \mathrm{d}t \right]\mathrm{d}t, \label{sol3}
\end{align}
where $\{c_{i}; \; i=1,\cdots,6\}$ are constants of integration. After taking integrations we arrive at
\begin{align}
u(t) &=c_{5}t+c_{6}, \label{sol11} \\
A(t) &=c_{3}+c_{7} \left(c_{5}t+c_{6} \right)^{1-2n_{j}}, \label{sol22} \\
w(t) &=c_{2}+c_{1}t+c_{8}t^{2}+c_{9}t^{3}+c_{10}\left(c_{5}t+c_{6} \right)^{1-2n_{j}}, \label{sol22}
\end{align}
where
\begin{align}
c_{7} &=\frac{c_{4}}{c_{5}(1-2n_{j})},\\
c_{8}&=\frac{c_{6}l_{2}}{2l_{3}},\\
c_{9}&=\frac{c_{5}l_{2}}{6l_{3}},\\
c_{10}&=\frac{l_{1} c^{2}_{4}}{2 n_{j} c^{2}_{5} l_{2} (2n_{j}-1)}.
\end{align}
Doing inverse transformations by the use of eqs. (\ref{phi12})--(\ref{bb123}) give solutions to our system:
\begin{align}
b_{1j,2j}(t)&=\left[c_{11}+c_{12}t+c_{13}t^2+c_{14}t^3+c_{15}t^4+c_{10}\left( c_{5}t+c_{6}\right)^{2-2n_{j}} \right]^{\frac{1}{m+2}}, \label{Rescalefactor}\\
\varphi_{1j,2j}(t)&=\frac{-\lambda}{2\mu}\ln\left[\left(c_{2}
+c_{1}t+c_{8}t^{2}+c_{9}t^{3} \right)(c_{5}t+c_{6})^{-1}+c_{10}(c_{5}t+c_{6})^{-2n_{j}} \right], \label{Rescalarfield}\\
A_{1j,2j}(t)&=c_{3}+c_{9} \left(c_{5}t+c_{6} \right)^{1-2n_{j}},  \label{Refour}
\end{align}
where
\begin{align}
c_{11}&=c_{1,2}c_{2}c_{6}(m+2),\\
c_{12}&=c_{1,2}(c_{2}c_{5}+c_{1}c_{6})(m+2),\\
c_{13}&=c_{1,2}(c_{1}c_{5}+c_{6}c_{8})(m+2),\\
c_{14}&=c_{1,2}(c_{5}c_{8}+c_{6}c_{9})(m+2),\\
c_{15}&=c_{1,2}c_{5}c_{9}(m+2).
\end{align}
In most of equations which contain the subscripts $1$, $2$, $3$, and $4$, these numbers refer to the type of class, for example, in eqs. (\ref{Rescalefactor})--(\ref{Refour}), the first number in each subscript refers to the type of class and the index $j$ refer to the type of the 4-vector potential $j=1$ (i.e. $k_{1}\neq 0$ and $k_{2}=0$) is for (\ref{AMU01}) and $j=2$ (i.e. $k_{1} = 0$ and $k_{2} \neq 0$) is for (\ref{AMU02}). The conserved currents $\{\mathbf{I}_{1}(\ref{CC1}),\mathbf{I}_{3j}(\ref{CC3})\}$ and $\{\mathbf{I}_{2}(\ref{CC2}),\mathbf{I}_{3j}(\ref{CC3})\}$ are carried by these solutions (i.e. $\{\mathbf{I}_{1}(\ref{CC1}),\mathbf{I}_{3j}(\ref{CC3})\}$ by $\{b_{1j}(t),\varphi_{1j}(t),A_{1j}(t)\}$; and $\{\mathbf{I}_{2}(\ref{CC2}),\mathbf{I}_{3j}(\ref{CC3})\}$ by $\{b_{2j}(t),\varphi_{2j}(t),A_{2j}(t)\}$). \\
In the section (\ref{data-sect}), these solutions are analyzed to demonstrate the most events of the universe evolution.\\
Note that because of $t^{2}$ term in the parentheses of eq.~(\ref{Rescalefactor}), the solutions obtained earlier in other papers for FRW background such as in refs. \cite{NN2}--\cite{NN3} are not recovered when the special case $m=1$ (namely FRW) is investigated, and consequently, other things are also different.\\

\noindent\rmybox{}{\subsection{The unusual cases: $\{c_{1}=c_{2}=1/2\}$ and $\{c_{1}=1/2$; $c_{2}= \; -1/2\}$ and CSSS-trick \label{sub-section2}}}\vspace{5mm}

In this sub-section, two cases $\{ c_{1}=c_{2}=1/2 \}$ (\textbf{C3}) and $\{ c_{1}=-c_{2}=1/2 \}$ (\textbf{C4}) are investigated. It is remembered that since $\mathbf{I}_{3}$ is carried automatically by field equations, hence we take $c_{4}=0$ in throughout this paper.\\
Regarding (\ref{liebracket}), if we want both $\mathbf{I}_{1}$ and $\mathbf{I}_{2}$ --- generated by $\mathbf{X}_{1}$ and $\mathbf{X}_{2}$, respectively --- are carried by field equations, a further symmetry does not produce. Hence, as mentioned in CSSS-trick, they may be combined in some suitable ways to lead to graceful results.\\
Now, by assuming that
\begin{align}\label{X-new1}
\mathbf{X}_{1+2}=\frac{1}{2}\left(\mathbf{X}_{1}+\mathbf{X}_{2} \right)
\; \Longrightarrow \; \mathbf{I}_{1+2}=\frac{1}{2}\left(\mathbf{I}_{1}+\mathbf{I}_{2} \right),
\end{align}
and
\begin{align}\label{X-new1}
\mathbf{X}_{1-2}=\frac{1}{2}\left(\mathbf{X}_{1}-\mathbf{X}_{2} \right)
\; \Longrightarrow \; \mathbf{I}_{1-2}=\frac{1}{2}\left(\mathbf{I}_{1}-\mathbf{I}_{2} \right),
\end{align}
are symmetries (symmetry generators) and conserved currents that are carried by \textbf{C3}-class and \textbf{C4}-class, respectively, we seek point transformations on the vector fields $\mathbf{X}_{1+2}$ and $\mathbf{X}_{1-2}$ for \textbf{C3}-class and \textbf{C4}-class respectively, such that
\begin{align}\label{X-new-equ1}
i_{\mathbf{X}_{1+2}} \mathrm{d}w=1, \quad i_{\mathbf{X}_{1+2}} \mathrm{d}u=0, \quad i_{\mathbf{X}_{1+2}} \mathrm{d}v=0
\end{align}
and
\begin{align}\label{X-new-equ-2}
i_{\mathbf{X}_{1-2}} \mathrm{d}w=1, \quad i_{\mathbf{X}_{1-2}} \mathrm{d}u=0, \quad i_{\mathbf{X}_{1-2}} \mathrm{d}v=0
\end{align}
whereas $i: \; (b,\varphi, A) \to (w,u,v)$ in which $w=w(b,\varphi, A)$, $u=u(b,\varphi, A)$, and $v=v(b,\varphi, A)$. In each case, $w$ is a cyclic variable.\\
Note that since we keep both $\mathbf{X}_{1}$ and $\mathbf{X}_{2}$, hence, two cyclic coordinates may practically be found by solving a system like
\begin{align}\label{hamin001}
i_{\mathbf{X}_{1}}\mathrm{d}Q^{1}=1, \quad
i_{\mathbf{X}_{2}}\mathrm{d}Q^{2}=1, \quad
i_{\mathbf{X}_{1}}\mathrm{d}Q^{i}=0\quad (i\neq 1), \quad
i_{\mathbf{X}_{2}}\mathrm{d}Q^{i}=0\quad (i\neq 2),
\end{align}
but, because both were combined in special ways and therefore, now, we have `one' (new/combined) symmetry for each class, then, for each class, (\ref{hamin001}) must be recast to
\begin{align}\label{hamin002}
i_{\mathbb{X}}\mathrm{d}Q^{1}=1, \qquad
i_{\mathbb{X}}\mathrm{d}Q^{i}=0\quad (i\neq 1),
\end{align}
where $\mathbb{X}$ is a mixed symmetry generator of $\mathbf{X}_{1}$ and $\mathbf{X}_{2}$. Therefore we have one cyclic variable for each class again. Indeed, (\ref{hamin001}) must be considered when one does not choose specified values for constants $c_{1}$ and $c_{2}$ and he wants two symmetries to be carried by the system independently and simultaneously.\\
In order to write down the equations and solutions of both classes in a unified (closed) forms, let us define some useful parameters:
\begin{align}
\theta &=\frac{c_{2}}{c_{1}}, \label{sppa1}\\
\delta_{1} &=\frac{1+\theta}{2},  \label{sppa2}\\
\delta_{2} &=\frac{1-\theta}{2}\label{sppa3}.
\end{align}
Therefore one has:
\begin{align}
\theta &\biggl|_{c_{1}=c_{2}=\frac{1}{2}} =+1, \qquad
\theta \biggl|_{c_{1}=-c_{2}=\frac{1}{2}} =-1, \label{sppa4}\\
\delta_{1} &\biggl|_{c_{1}=c_{2}=\frac{1}{2}}=+1,  \qquad
\delta_{1} \biggl|_{c_{1}=-c_{2}=\frac{1}{2}}=0,  \label{sppa5}\\
\delta_{2} &\biggl|_{c_{1}=c_{2}=\frac{1}{2}}=0,  \qquad
\delta_{2} \biggl|_{c_{1}=-c_{2}=\frac{1}{2}}=+1  \label{sppa6}.
\end{align}
With these definitions at hand, the solutions (\ref{sol for NS1})--(\ref{sol for NS4}) for both cases are now written in unified forms:
\begin{align}
\gamma &=-\sqrt{4m+2} \; b^{-(m+2)/2} \left[\delta_{1} \sinh(\mu \varphi)+\delta_{2} \cosh(\mu \varphi) \right],   \label{1sppa}\\
\beta &=b^{-m/2} \left[\delta_{1} \cosh(\mu \varphi)+\delta_{2} \sinh(\mu \varphi) \right], \label{2sppa}\\
\alpha &=0,\label{3sppa}\\
V(\varphi) &=V_{0} \left[\delta_{1} \sinh^{2}(\mu \varphi) +\delta_{2} \cosh^{2}(\mu \varphi) \right],   \label{11sppa}\\
f(\varphi) &=f_{0} \left[\delta_{1} \sinh^{n_{j}}(\mu \varphi) +\delta_{2} \cosh^{n_{j}}(\mu \varphi) \right]^{n_{j}/2},  \label{12sppa}
\end{align}
and the symmetry generators and conserved currents turn also out to be
\begin{align}
\mathbb{X}= \; & b^{-m/2} \left[\delta_{1} \cosh(\mu \varphi)+\delta_{2} \sinh(\mu \varphi) \right]\frac{\partial}{\partial b} \nonumber \\
&-\sqrt{4m+2} \; b^{-(m+2)/2}\left[\delta_{1} \sinh(\mu \varphi)+\delta_{2} \cosh(\mu \varphi) \right]\frac{\partial}{\partial \varphi},   \label{13sppa}\\
\mathbb{I}= \; & 2(2m+1) b^{m/2} \left[\delta_{1} \cosh(\mu \varphi)+\delta_{2} \sinh(\mu \varphi) \right]\dot{b} \nonumber \\
& +\sqrt{4m+2}\; b^{(m+2)/2}\left[\delta_{1} \sinh(\mu \varphi)+\delta_{2} \cosh(\mu \varphi) \right]\dot{\varphi}.  \label{14sppa}
\end{align}
One of the fruits of our different taken procedure is cleared here: Further interesting forms of the potentials namely $V_{\mathrm{\textbf{C3}}}(\varphi)=V_{0} \sinh^{2}(\mu \varphi)$ and $V_{\mathrm{\textbf{C4}}}(\varphi)=V_{0} \cosh^{2}(\mu \varphi)$ were acquired.
In section (\ref{data-sect}), some interesting discussions about the obtained forms of potentials are performed.\\
Solving eqs. (\ref{X-new-equ1}) and (\ref{X-new-equ-2}) leads to
\begin{align}
w &=\left(\frac{2\theta}{m+2} \right) b^{(m+2)/2} \left[\delta_{1} \cosh(\mu \varphi)+\delta_{2} \sinh(\mu \varphi) \right],   \label{4sppa}\\
u &=\left(\frac{2\theta}{m+2} \right) b^{(m+2)/2} \left[\delta_{1} \sinh(\mu \varphi)+\delta_{2} \cosh(\mu \varphi) \right],  \label{6sppa}\\
v &=A.   \label{7sppa}\\
\end{align}
So, the corresponding inverse transformations are
\begin{align}
\varphi &= \frac{1}{\mu}\mathrm{arctanh} \left(\frac{\delta_{1}u+\delta_{2}w}{\delta_{1}w+\delta_{2}u} \right)=\frac{1}{2\mu} \ln \left(\frac{w+u}{\theta w-\theta u} \right),   \label{7sppa}\\
b &= \left(\frac{m+2}{2} \right)^{2/(m+2)} \left(\frac{w^2 - u^2}{\theta} \right)^{1/(m+2)},  \label{8sppa}\\
A &=v. \label{25808sppa}
\end{align}
Therefore, the potential, (\ref{11sppa}), and coupling function, (\ref{12sppa}), would be
\begin{align}
V(u,w) &=V_{0} \left(\frac{\theta \; u^2}{w^2 - u^2} \right),   \label{9sppa}\\
f(u,w) &=f_{0} \left(\frac{\theta \; u^2}{w^2 - u^2} \right)^{n_{j}/2}.  \label{10sppa}
\end{align}
Again, it must be noted that the coordinate transformation is not unique, but our choices are very advantageous. Now, like the previous cases in sub-section (\ref{sub-section1}), the point-like Lagrangian (\ref{point-like(LRS)}) is split into two Lagrangians for each class and they may be rewritten in terms of cyclic variables as
\begin{align}\label{15sppa}
\mathcal{L}_{3j,4j}=l_{4} u^{2}+\theta l_{5} \left(\dot{w}^2 - \dot{u}^2 \right)+l_{6j} u^{2n_{j}} \dot{A}^{2},
\end{align}
where:
\begin{align}\label{da7}
l_{4} &=\frac{V_{0}}{4} (m+2)^2, \nonumber \\
l_{5} &=2m+1,  \nonumber \\
l_{6j} &= \frac{-k^{2}_{j}}{2^{2n_{j}+1}} (m+2)^{2n_{j}} f^{2}_{0}.
\end{align}
Indeed, (\ref{15sppa}) represents four types of different Lagrangians (i.e $\mathcal{L}_{31}$, $\mathcal{L}_{32}$, $\mathcal{L}_{41}$, and $\mathcal{L}_{42}$).\\
The Lagrangians (\ref{15sppa}) lead to the following field equations with respect to $u$, $w$, and $A$, respectively:
\begin{align}
&\ddot{w}=0, \label{da8.1} \\
&\theta l_{5} \ddot{u} +2n_{j} l_{6j} u^{2n_{j}-1}\dot{A}^2+l_{4}u=0, \label{da8.2} \\
&4n_{j} l_{6j} u^{2n_{j}-1} \dot{u} \dot{A}+2 l_{6j} u^{2n_{j}} \ddot{A}=0. \label{da8.3}
\end{align}
Again, in these cases, the conserved currents (\ref{14sppa}) in terms of the cyclic variables do not add new equations to our systems. The above system yields the following solutions:
\begin{align}
w(t)&=c_{20}t+c_{21} \label{9.1} \\
u(t)&=\left( \; c_{24}\sinh \left[c_{25}(t+c_{23}) \right] \; \right)^{1/(1+n_{j})}, \label{9.2} \\
A_{3j,4j}(t)&= c_{26} \int \left( \sinh \left[c_{25}(t+c_{23}) \right] \right)^{\frac{-2n_{j}}{1+n_{j}}} \mathrm{d}t ,\label{9.3}
\end{align}
in which
\begin{align}\label{10.1}
c_{24} &= \left(\frac{- \theta l_{5} \; c^{2}_{22}}{2l_{4}l_{6j}} \right)^{\frac{1+n_{j}}{n_{j}}},\\
c_{25} &=\frac{(1+n_{j})^{2}}{- \theta} \; \frac{l_{4}}{l_{5}},\\
c_{26} &= \left(\frac{c_{22}}{2 l_{6j}} \right) c^{\frac{-2n_{j}}{1+n_{j}}}_{24}.
\end{align}
Finally, performing inverse transformations, the following solutions are acquired:
\begin{align}\label{solC3-C4-1}
\varphi_{3j,4j}=\frac{1}{2 \mu}\ln \left[\frac{c_{20}t+c_{21}+\left( c_{24}\sinh [c_{25}(t+c_{23})]\right)^{1/(1+n_{j})}}{\theta c_{20}t+\theta c_{21}-\theta \left( c_{24}\sinh [c_{25}(t+c_{23})]\right)^{1/(1+n_{j})}} \right],
\end{align}
\begin{align}\label{solC3-C4-2}
b_{3j,4j}=\left(\frac{m+2}{2} \right)^{2/(m+2)} \left[\frac{\left(c_{20}t+c_{21} \right)^2 - \left(c_{24}\sinh [c_{25}(t+c_{23})] \right)^{2/(1+n_{j})}}{\theta} \right]^{1/(m+2)}.
\end{align}
The forms of the 4-vector potentials are given by (\ref{9.3}). When the value of $n_{j}$ is not exactly clear, taking this integral is somewhat complicated, hence we kept it in the form (\ref{9.3}).\\
In section (\ref{data-sect}), by singling suitable values of constant parameters out, the analysis of all solutions are easily carried out. The conserved currents $\{\mathbf{I}_{3j},\mathbb{I}=(\mathbf{I}_{1}+\mathbf{I}_{2})/2\}$ and $\{\mathbf{I}_{3j},\mathbb{I}=(\mathbf{I}_{1}-\mathbf{I}_{2})/2\}$ are carried by $\{b_{3j},\varphi_{3j},A_{3j}\}$ (\textbf{C3}-class) and $\{b_{4j},\varphi_{4j},A_{4j}\}$ (\textbf{C4}-class), respectively.\\

\noindent\pqrmybox{red}{\textbf{$\bullet$} \textbf{An important point.}}\\

Perhaps, it seems that there are degeneracies in the solutions of all cases studied in this chapter, due to the existence of $n_{j}$, when one has $n_{1}=n_{2}$, but it is completely wrong idea because $n_{1}=n_{2}$ holds only for $m=1$, namely FRW case. On the other hand, in FRW case, we do not have permission to adopt one of the 4-vector potentials (\ref{AMU01}) and (\ref{AMU02}) because of the background geometry. Indeed, in FRW space-time, we must take the 4-vector potential of the form $A_{\mu}=\left( \chi(t); k_{1}A(t),k_{1}A(t),k_{1}A(t)\right)$ (i.e. we must take $k_{1}=k_{2}/\sqrt{2}$) since both (\ref{AMU01}) and (\ref{AMU02}) violate the cosmological principle on which FRW metric is based.\\

\noindent\hrmybox{}{\section{Satisfaction of Maxwell's equations \label{Max-sect}}}\vspace{5mm}

In this section, demonstrating the satisfaction of Maxwell's equations is our objective.\\
Maxwell's equations in curved space-time in terms of the components of the field tensor $\mathbf{F}$ are \cite{MTW}
\begin{align}
&F_{\alpha \beta, \gamma}+F_{\beta \gamma, \alpha}+F_{\gamma \alpha, \beta}=0,\label{ME1} \\
&{F^{\alpha \beta}}_{,\beta}=-4 \pi J^{\alpha}; \nonumber \\ & \left\{\begin{array}{ll} \text{if }\; \alpha = 0: &{}  \; J^0 = \rho = \text{charge density,} \\ \text{if }\; \alpha \ne 0: &{} \; (J^1, J^2, J^3)\\ &{}= \text {components of current density,} \end{array} \right. \label{ME2}
\end{align}
where \(\{J^{\alpha }\); \(\alpha \in \{0, 1, 2, 3\}\}\) are the components of the 4-current \(\mathbf {J}\). The usual forms of Maxwell's equations may easily be acquired because eq. (\ref{ME1}) reduces to \(\mathbf {\nabla } \mathbf {\cdot } \mathbf {B} = 0\) when we take \(\alpha = 1\), \(\beta =2\), and \(\gamma = 3\); and it reduces to \(\partial \mathbf {B} / \partial t + \mathbf {\nabla } \times \mathbf {E} = 0\) when one puts any index, e.g., \(\alpha = 0\), and through eq. (\ref{ME2}), two of Maxwell's equations,
\(\mathbf {\nabla \cdot E }= 4 \pi \rho \) (the electrostatic equation), and \(\partial \mathbf {E} / \partial t - \mathbf {\nabla } \times \mathbf {B} = -4 \pi \mathbf {J}\) (the electrodynamic equation), are obtained by setting \(\alpha = 0\) and \(\alpha \ne 0\), respectively. Therefore, it may be claimed that through eq. (\ref{ME1}) magnetodynamics and magnetostatics, and through eq. (\ref{ME2}) electrodynamics and electrostatics have been unified in one geometric law.\\
Regarding the 4-current of our case of study which is \(\mathbf {J}=(J^0, J^1, J^2, J^3) = (0, 0, 0, 0)\), eqs. (\ref{ME1}) and (\ref{ME2}) for the electromagnetism part of the action (\ref{action}), i.e.
\begin{align}\label{EMPA}
\mathcal{L}_{\mathrm{EM}}= &{}-\frac{1}{4}\int \mathrm{d}^4 x \sqrt{-g}f^2(\varphi ) F_{\mu \nu }F^{\mu \nu } \nonumber \\= &{} -\frac{1}{4}\int \mathrm{d}^4 x \sqrt{-g} g^{\alpha \beta } g^{\mu \nu} f^2(\varphi) F_{\mu \alpha} F_{\nu \beta},
\end{align}
turn out to be
\begin{align}\label{EM3}
&\partial ^{\alpha} \left( \sqrt{-g} f^2 F^{\beta \gamma} \right)+ \partial ^{\beta} \left( \sqrt{-g} f^2 F^{\gamma \alpha} \right)\nonumber \\&\quad +\;\partial ^{\gamma} \left( \sqrt{-g} f^2 F^{\alpha \beta} \right) =0,
\end{align}
and
\begin{align}\label{EM4}
&\partial_{\mu} \left[ \sqrt{-g}g^{\alpha \beta} g^{\mu \nu} f^2 F_{\nu \beta}\right] =0 \nonumber \\&\quad \longrightarrow \quad \left( \sqrt{-g}f^2 F^{\alpha \mu } \right)_{,\mu } = 0,
\end{align}
respectively. In our case, eqs. (\ref{EM3}) and (\ref{EM4}) may be recast the same equation, viz.
\begin{align} \label{EM5}
\frac{\partial}{\partial t} \left(k_{j}b^{(2+m)n_{j}} f^2 \dot{A} \right)=0.
\end{align}
In general, there is an Einstein summation convention over the subscript $j$, nonetheless, it also holds true for each of indices $j=1$ ($k_{1} \neq 0$, $k_{2}=0$) and $j=2$ ($k_{1}=0$, $k_{2} \neq 0$). According to (\ref{EM5}), \( k_{j}b^{(2+m)n_{j}} f^2 \dot{A} \) is a time-independent term, so it is a constant of motion, as it emerged by the use of Noether symmetry approach (See (\ref{CC3}); \( \mathbf{I}_{3}=k_{j}b^{(2+m)n_{j}} f^2 \dot{A} \)). As we observe, eq. (\ref{EM5}) is equivalent to the third field equation namely eq. (\ref{4v-equation}) and hence Maxwell's equations are satisfied automatically. And also it is needless to consider $\mathbf{X}_{3}$ (or $\mathbf{I}_{3}$) because $\mathbf{I}_{3}$ is carried automatically.\\
Before terminating this section, let us define the electric \(\mathbf {E}\) and magnetic \(\mathbf {B}\) fields covariantly, which are seen by an observer who is characterized by the 4-velocity vector \(u^{\mu }\). For the components of these fields, one has \cite{JDBARROW}
\begin{align}\label{EB-def}
E_{\mu } = u^{\nu } F_{\mu \nu}, \quad B_{\mu } = \frac{1}{2} \varepsilon_{\mu \nu \kappa} F^{\nu \kappa},
\end{align}
where the tensor \(\varepsilon_{\mu \nu \kappa }\) is defined by the relation
\begin{align}\label{EB-def2}
\varepsilon_{\mu \nu \kappa} = \eta_{\mu \nu \kappa \lambda} u^{\lambda},
\end{align}
in which \(\eta_{\mu \nu \kappa \lambda }\) is an antisymmetric permutation tensor of spacetime with \(\eta ^{0123} = 1/\sqrt{-g}\) or \(\eta_{0123} = \sqrt{-g}\). In cosmic time for a comoving observer with \(u^{\mu }= (1, 0, 0, 0)\), we obtain
\begin{align}
E_{\mu }= & {} \left\{ \begin{array}{ll} E_{i} = - \dot{A}_{i}; &{} \quad \hbox {for} \; \mu = i =1, 2, 3 \\ 0; &{} \quad \hbox {for} \; \mu = 0 \end{array}\right. , \label{EB-def3} \\
B_{\mu }= & {} \left\{ \begin{array}{ll} B_{i} = \frac{1}{a} \epsilon_{ijk} \partial_{j} A_{k}; &{} \quad \hbox {for} \; \mu =i, j, k \in \{1, 2, 3\} \\ 0; &{} \quad \hbox  {otherwise,} \end{array} \right. \label{EB-def4}
\end{align}
where \(\epsilon_{ijk}\) is the well-known Levi-Civita symbol with \(\epsilon_{123}=1\). Therefore, after specifying the form of the 4-vector potential, forms of the electric and magnetic fields would be achievable.\\

\noindent\hrmybox{}{\section{Data Analysis \label{data-sect}}}\vspace{5mm}

\begin{figure*}
\centering
\includegraphics[width=6.1 in, height=2.55 in]{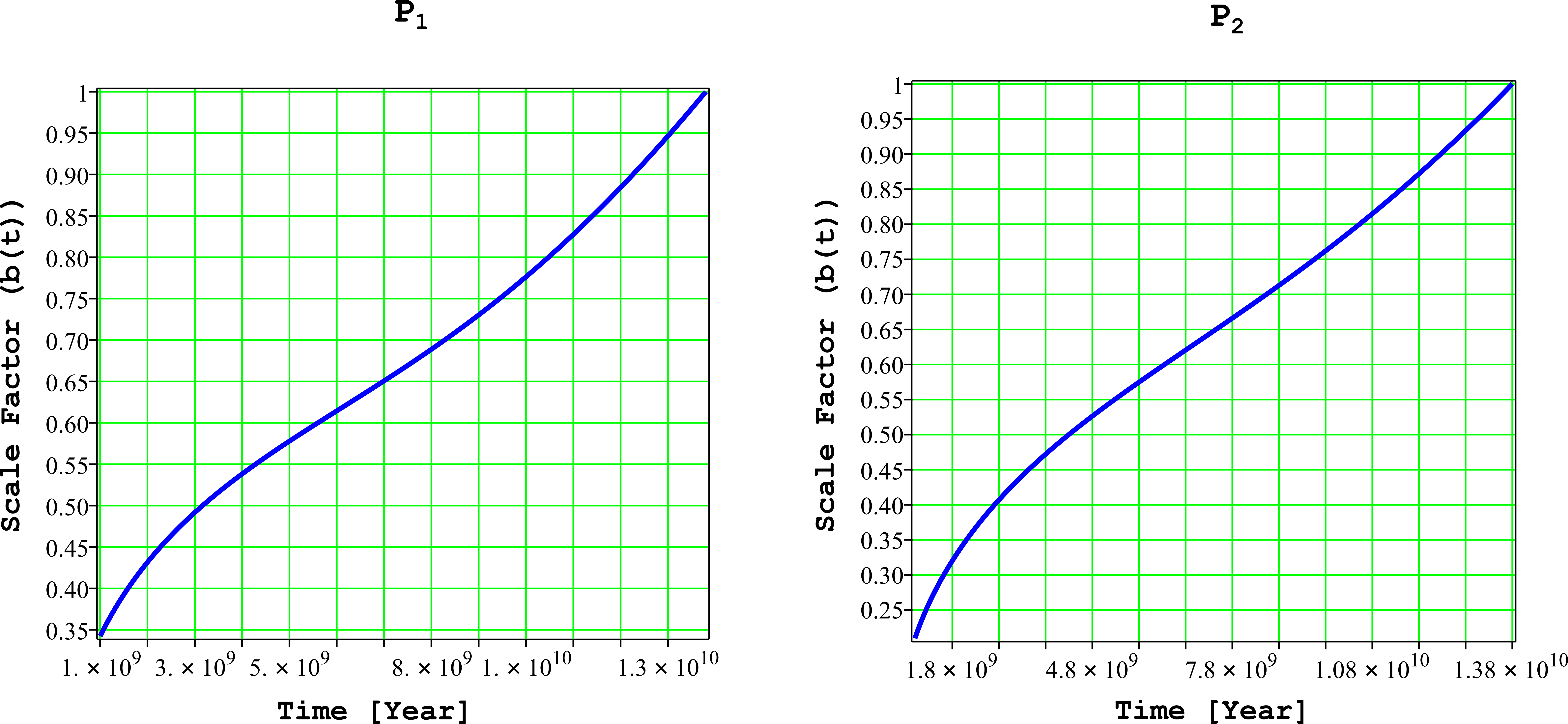}\\
\caption{Plots $\texttt{P}_{1}$ and $\texttt{P}_{2}$ indicate the scale factor versus time at the time interval $[1\mathrm{Gyr}, 13.801 \mathrm{Gyr}]$ for \textbf{C1} and \textbf{C4} classes, respectively.}\label{0a-t}
\end{figure*}
\begin{figure*}
\centering
\includegraphics[width=6.2 in, height=2.55 in]{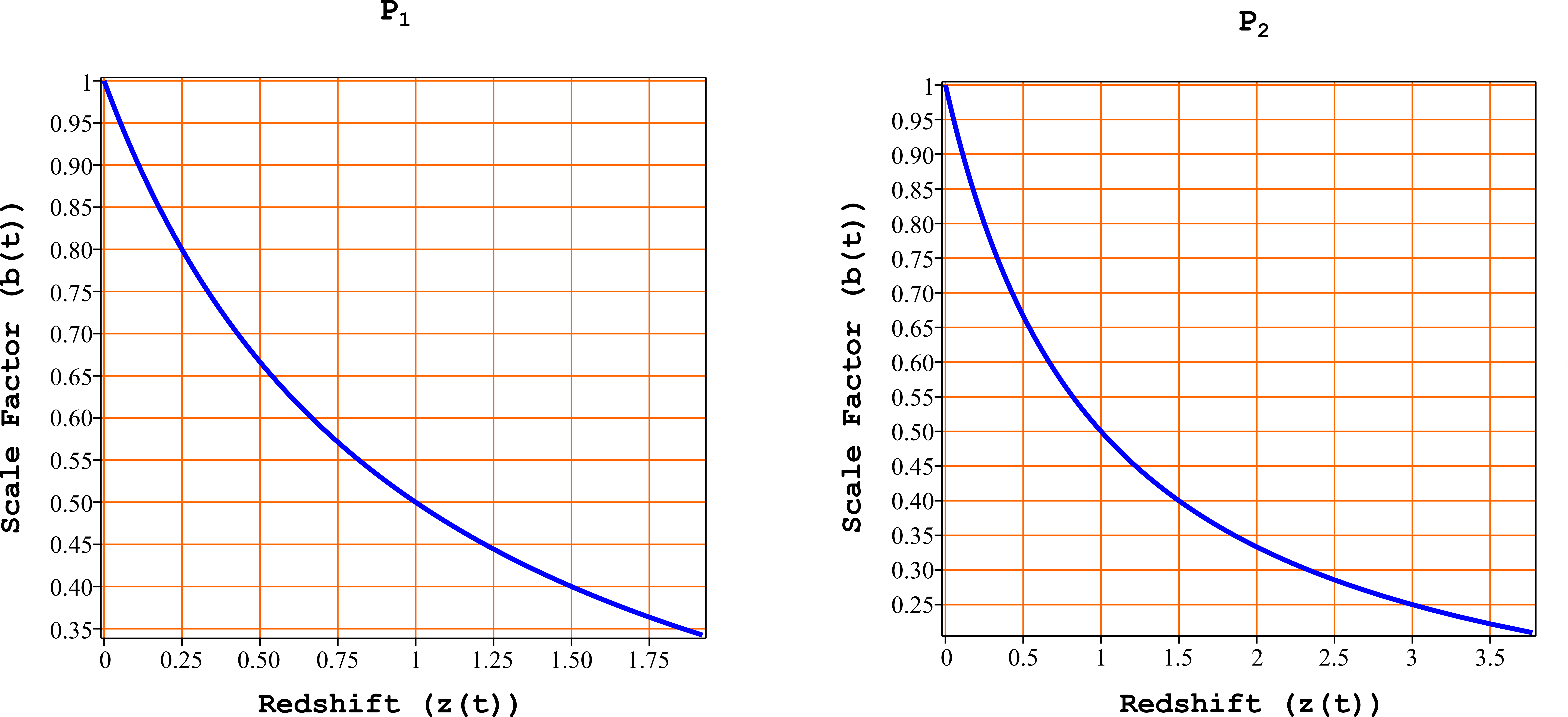}\\
\caption{Plots $\texttt{P}_{1}$ and $\texttt{P}_{2}$ show the scale factor versus redshift at the time range $[1\mathrm{G yr}, 13.801 \mathrm{G yr}]$ for \textbf{C1} and \textbf{C4} classes, respectively.}\label{0a-z}
\end{figure*}
\begin{figure*}
\centering
\includegraphics[width=6.2 in, height=2.55 in]{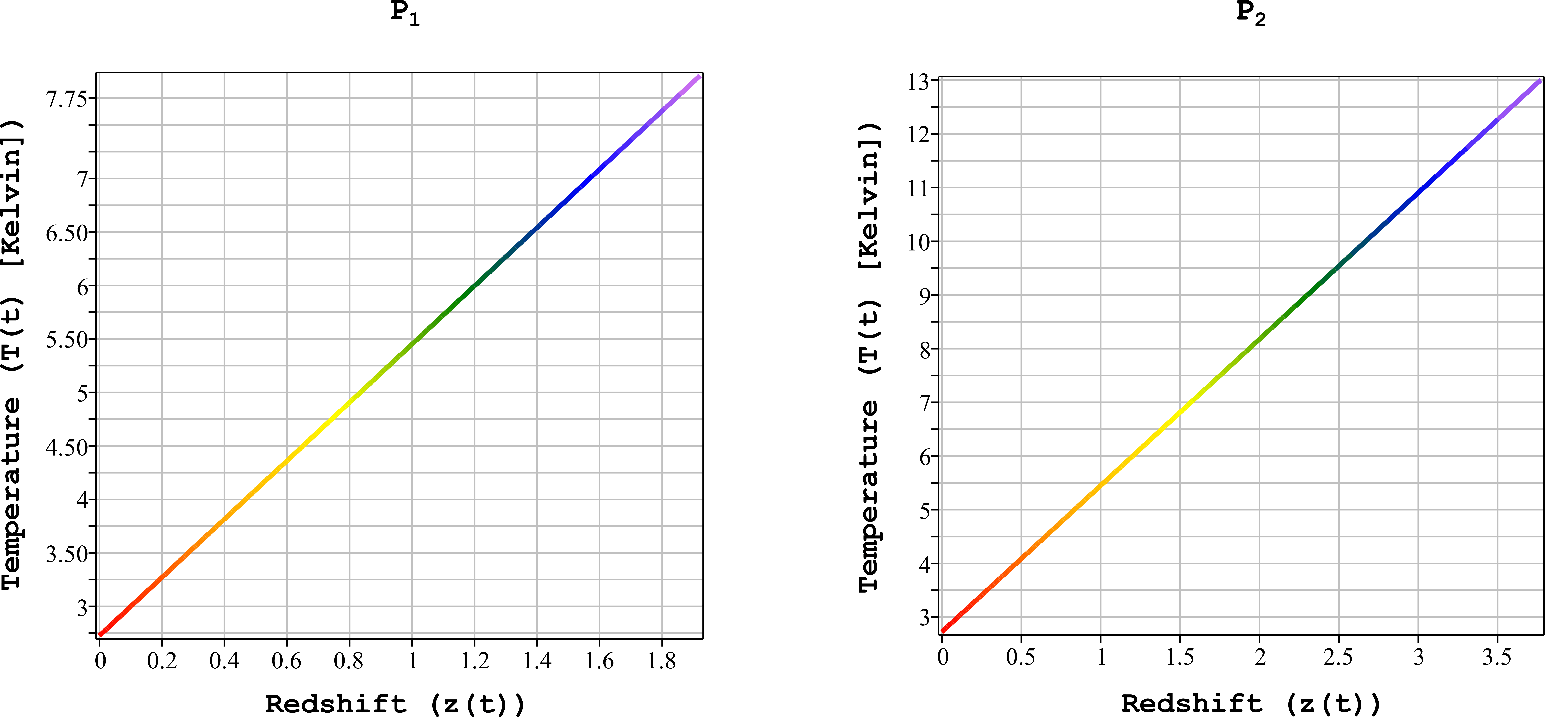}\\
\caption{Plots $\texttt{P}_{1}$ and $\texttt{P}_{2}$ demonstrate the temperature versus redshift at the time interval $[1\mathrm{G yr}, 13.801 \mathrm{G yr}]$ for \textbf{C1} and \textbf{C4} classes, respectively.}\label{0T-z}
\end{figure*}
\begin{figure*}
\centering
\includegraphics[width=6.2 in, height=2.55 in]{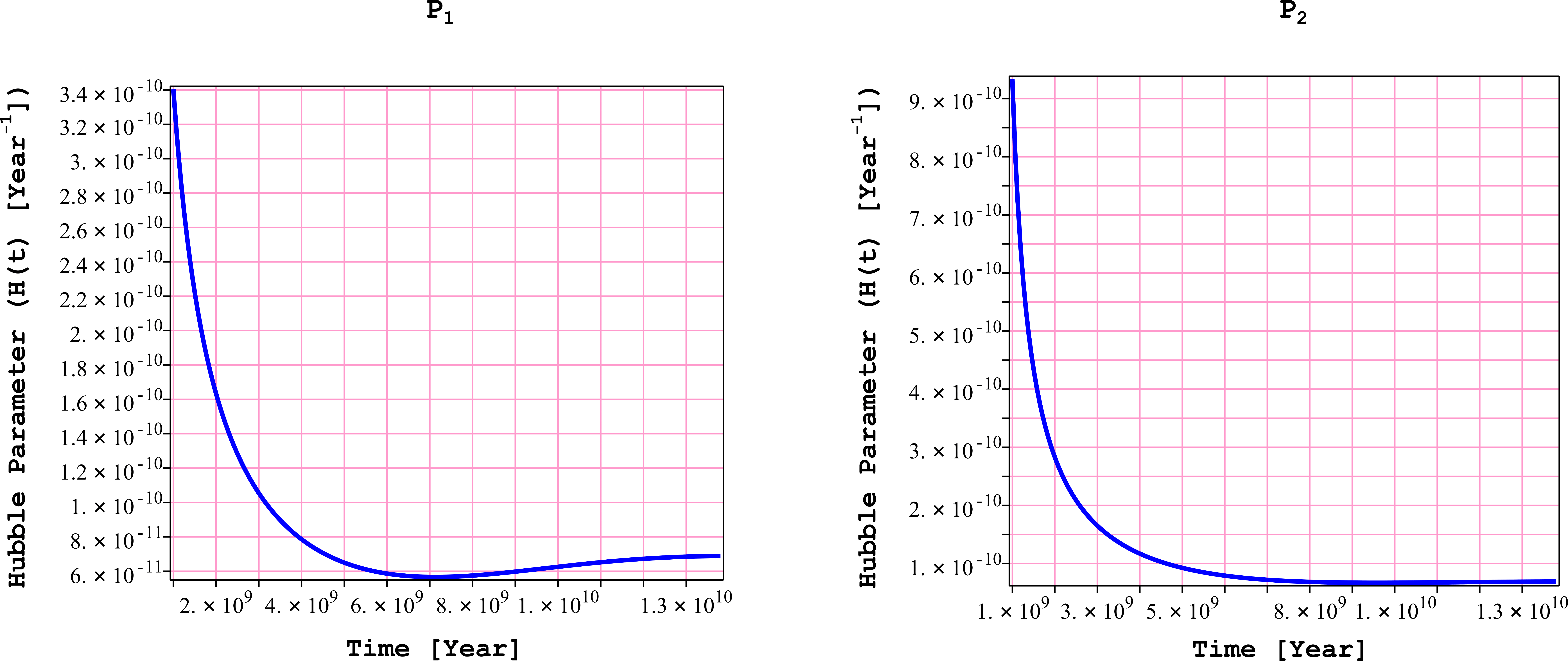}\\
\caption{Plots $\texttt{P}_{1}$ and $\texttt{P}_{2}$ indicate the Hubble parameter versus time at the time range $[1\mathrm{G yr}, 13.801 \mathrm{G yr}]$ for \textbf{C1} and \textbf{C4} classes, respectively.}\label{0H-t}
\end{figure*}
\begin{figure*}
\centering
\includegraphics[width=6.2 in, height=2.55 in]{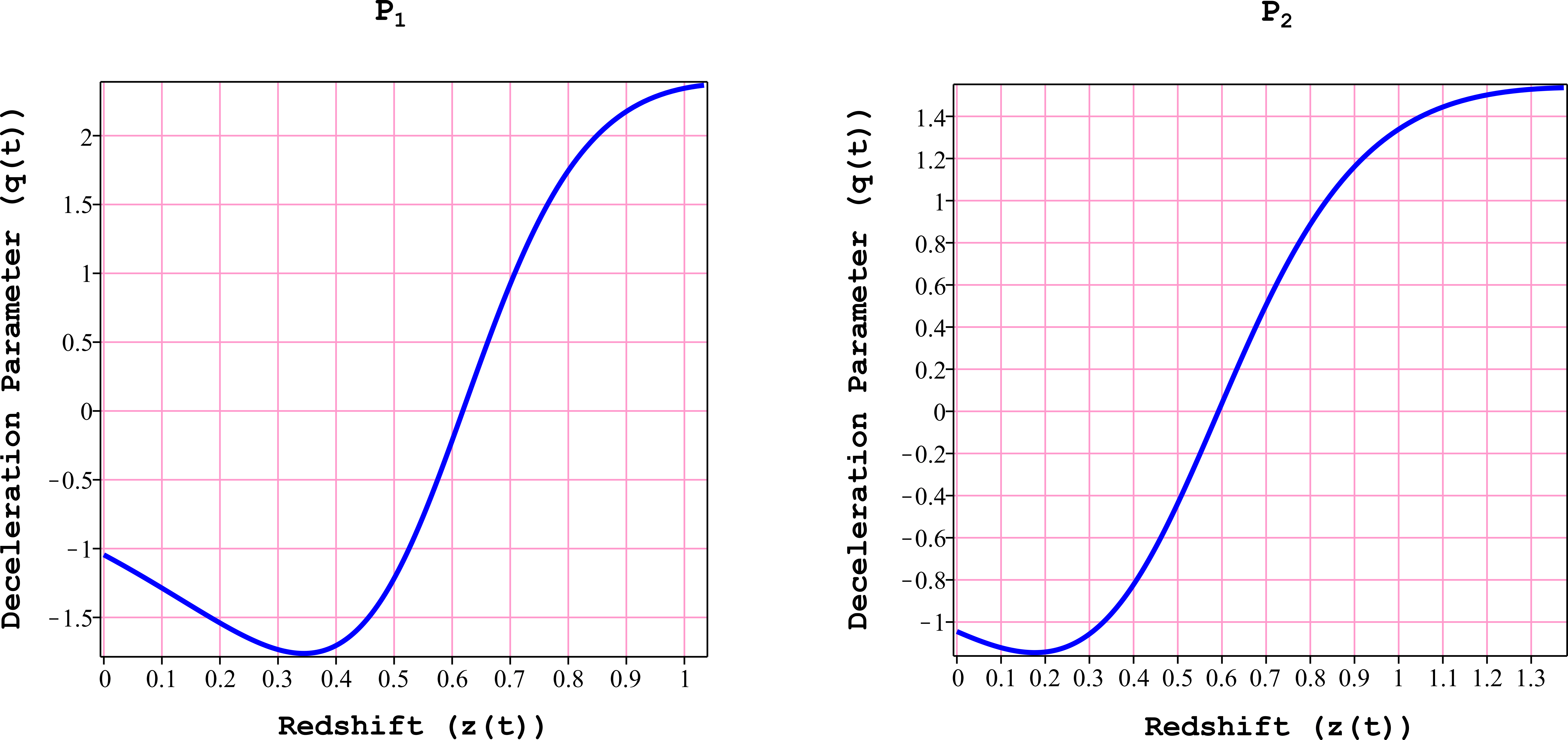}\\
\caption{Plots $\texttt{P}_{1}$ and $\texttt{P}_{2}$ show the deceleration parameter versus redshift at the time interval $[3\mathrm{G yr}, 13.801 \mathrm{G yr}]$ for the \textbf{C1} and \textbf{C4} classes, respectively.}\label{0q-z}
\end{figure*}
\begin{figure*}
\centering
\includegraphics[width=6.2 in, height=2.55 in]{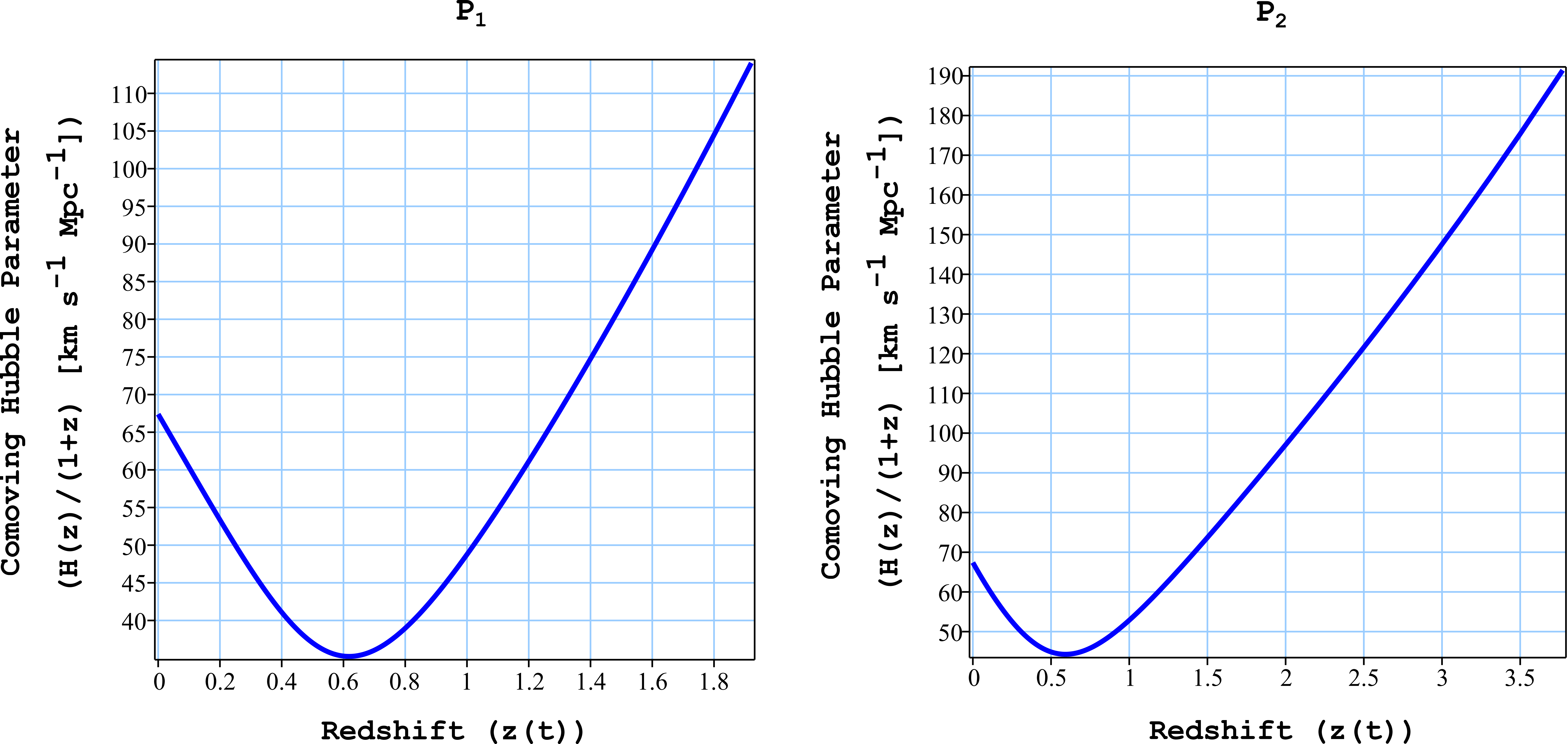}\\
\caption{Plots $\texttt{P}_{1}$ and $\texttt{P}_{2}$ demonstrate the comoving Hubble parameter, $H(z)/(1+z)$, versus redshift at the time range $[1\mathrm{G yr}, 13.801 \mathrm{G yr}]$ for \textbf{C1} and \textbf{C4} classes, respectively.}\label{0H-z}
\end{figure*}
\begin{figure*}
\centering
\includegraphics[width=6.2 in, height=2.55 in]{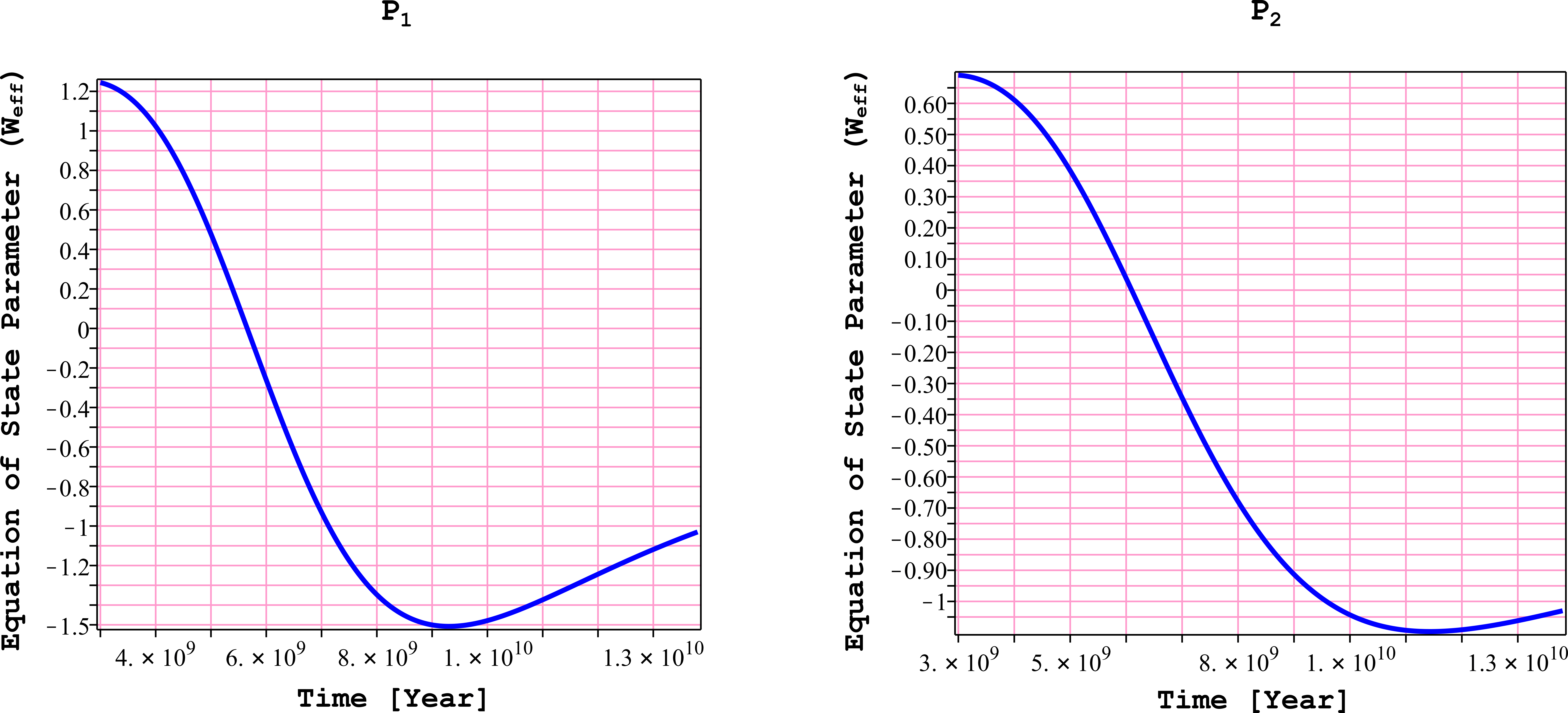}\\
\caption{Plots $\texttt{P}_{1}$ and $\texttt{P}_{2}$ indicate the equation of state parameter versus time at the time interval $[3\mathrm{G yr}, 13.801 \mathrm{G yr}]$ for \textbf{C1} and \textbf{C4} classes, respectively.}\label{0W-t}
\end{figure*}
\begin{figure*}
\centering
\includegraphics[width=6.2 in, height=2.55 in]{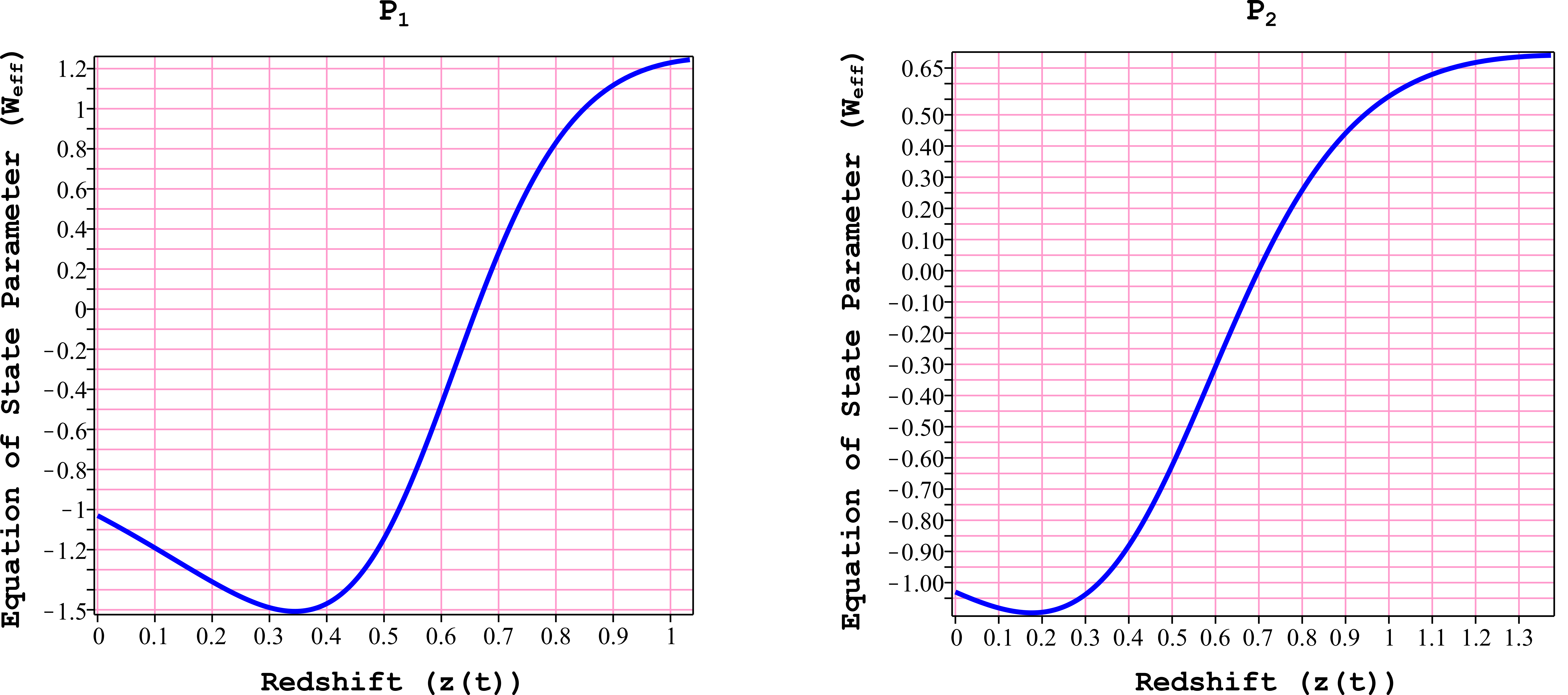}\\
\caption{Plots $\texttt{P}_{1}$ and $\texttt{P}_{2}$ show the equation of state parameter versus redshift at the time range $[3\mathrm{G yr}, 13.801 \mathrm{G yr}]$ for \textbf{C1} and \textbf{C4} classes, respectively.}\label{0W-z}
\end{figure*}
\begin{figure*}
\centering
\includegraphics[width=6.2 in, height=2.55 in]{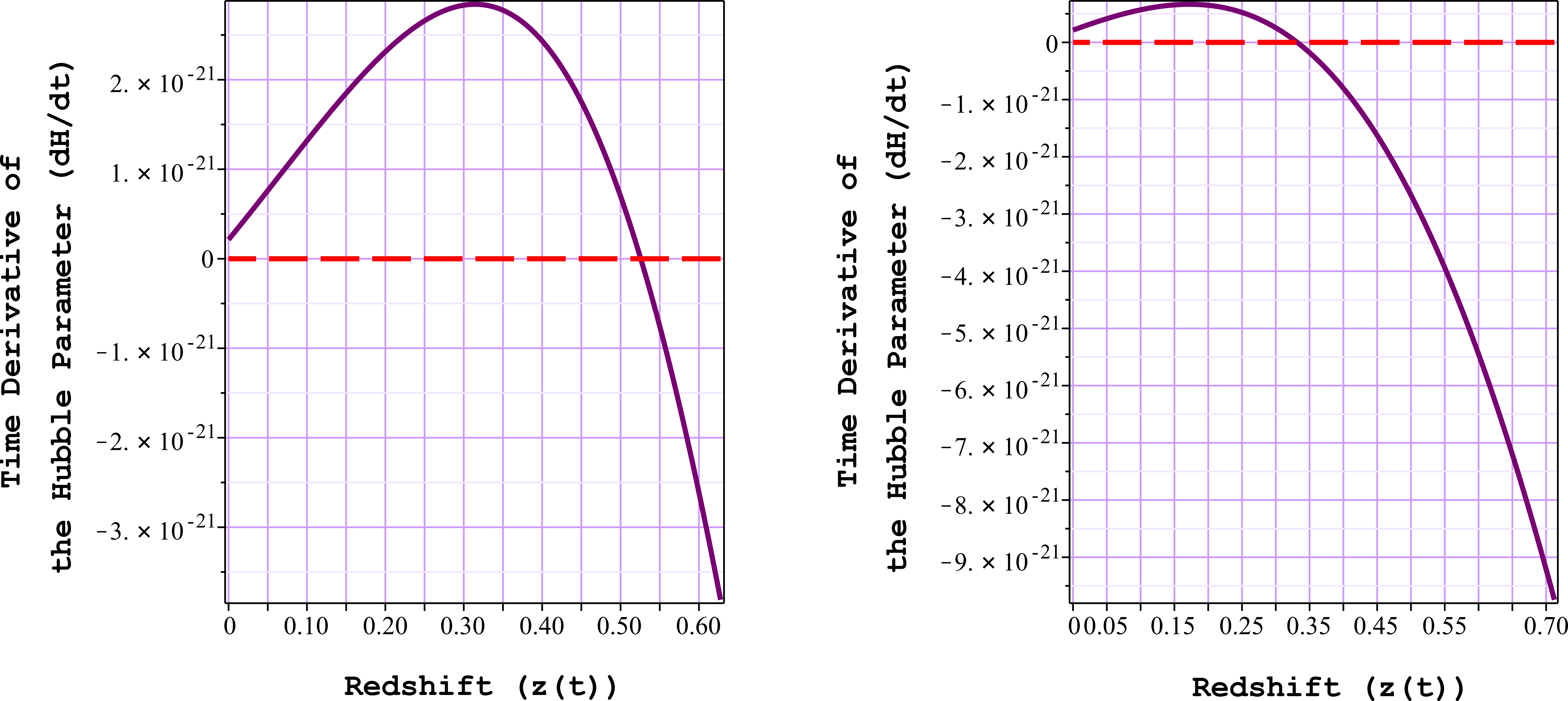}\\
\caption{Plots $\texttt{P}_{1}$ and $\texttt{P}_{2}$ demonstrate $\mathrm{d}H/\mathrm{d}t$ versus redshift at the time interval $[6\mathrm{G yr}, 13.801 \mathrm{G yr}]$ for \textbf{C1} and \textbf{C4} classes, respectively.}\label{0dH-z}
\end{figure*}
\begin{figure*}
\centering
\includegraphics[width=6.1 in, height=2.55 in]{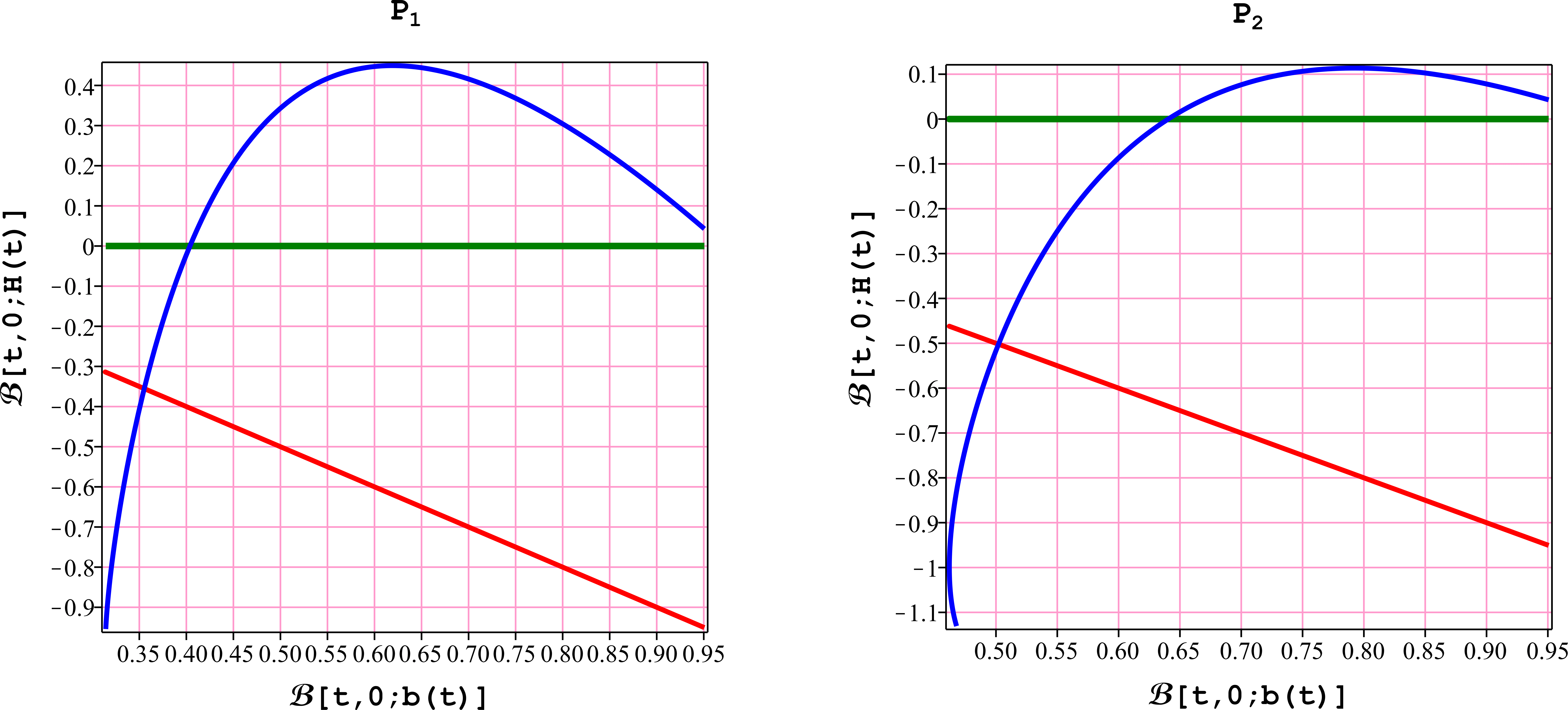}\\
\caption{Plots $\texttt{P}_{1}$ and $\texttt{P}_{2}$ indicate $\mathfrak{B}[t;0,H(t)]$ with respect to $\mathfrak{B}[t;0,b(t)]$ at the time range $[4\mathrm{G yr}, 13.801 \mathrm{G yr}]$ for \textbf{C1} and \textbf{C4} classes, respectively.}\label{0B-function}
\end{figure*}
\begin{figure*}
\centering
\includegraphics[width=6.2 in, height=2.55 in]{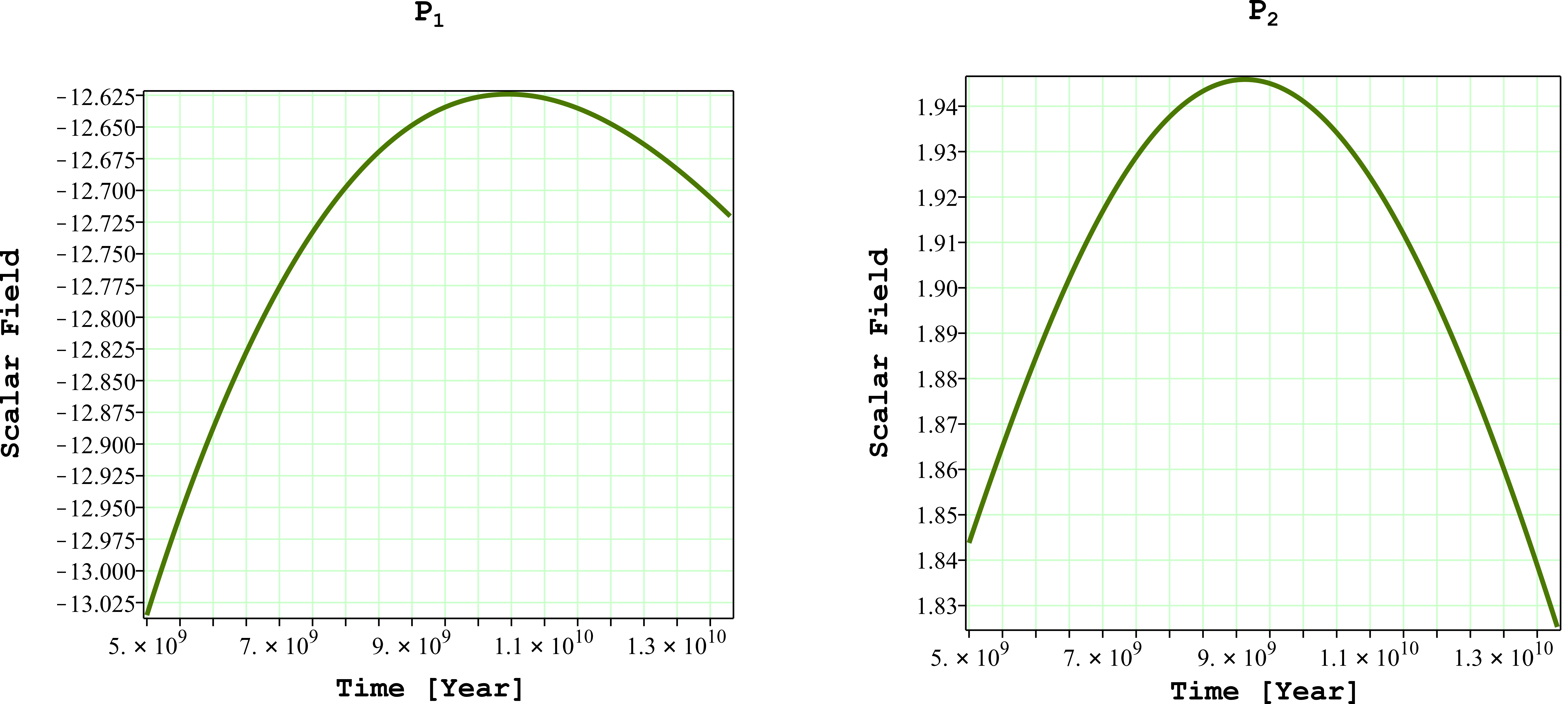}\\
\caption{Plots $\texttt{P}_{1}$ and $\texttt{P}_{2}$ show the scalar field versus time at the time interval $[5\mathrm{G yr}, 13.801 \mathrm{G yr}]$ for \textbf{C1} and \textbf{C4} classes, respectively.}\label{0p-t}
\end{figure*}
\begin{figure*}
\centering
\includegraphics[width=6.2 in, height=2.55 in]{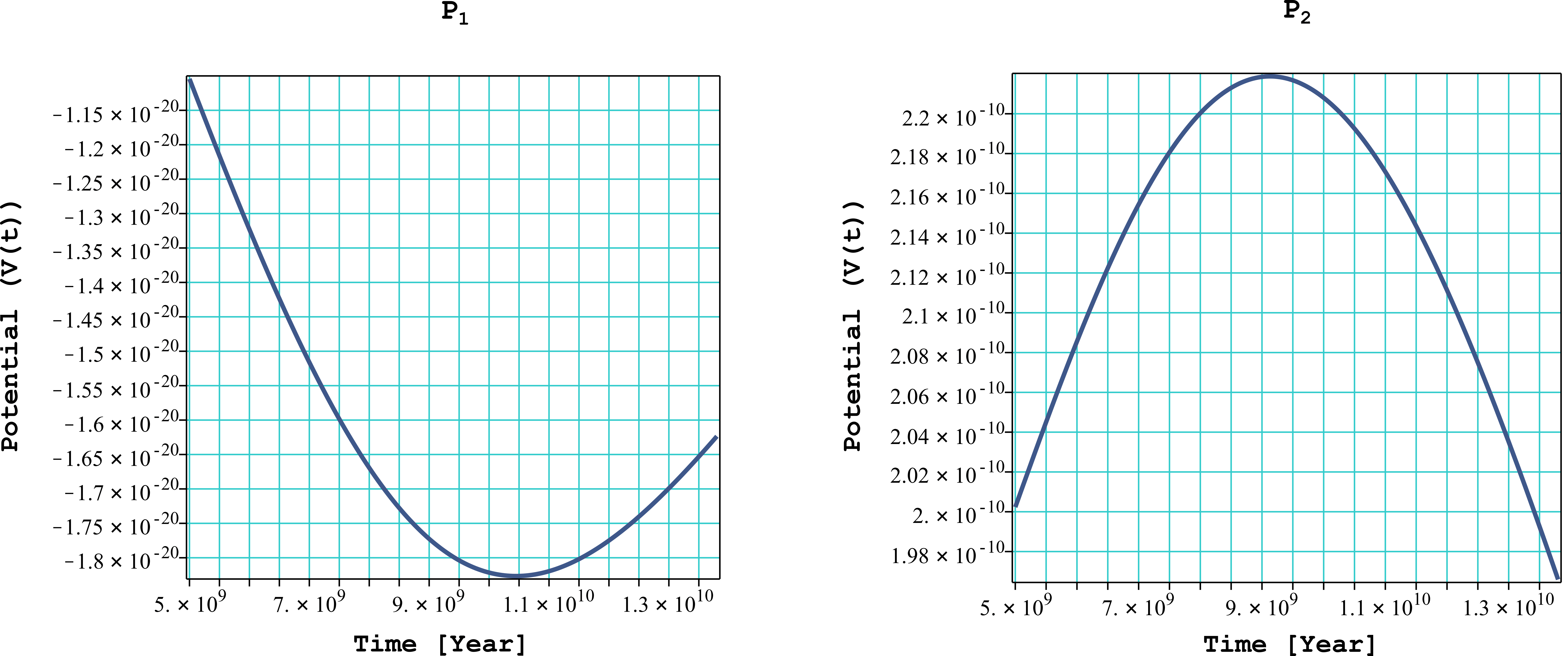}\\
\caption{Plots $\texttt{P}_{1}$ and $\texttt{P}_{2}$ demonstrate the scalar potential, $V$, versus time at the time range $[5\mathrm{G yr}, 13.801 \mathrm{G yr}]$ for \textbf{C1} and \textbf{C4} classes, respectively.}\label{0V-t}
\end{figure*}
\begin{figure*}
\centering
\includegraphics[width=6.2 in, height=2.55 in]{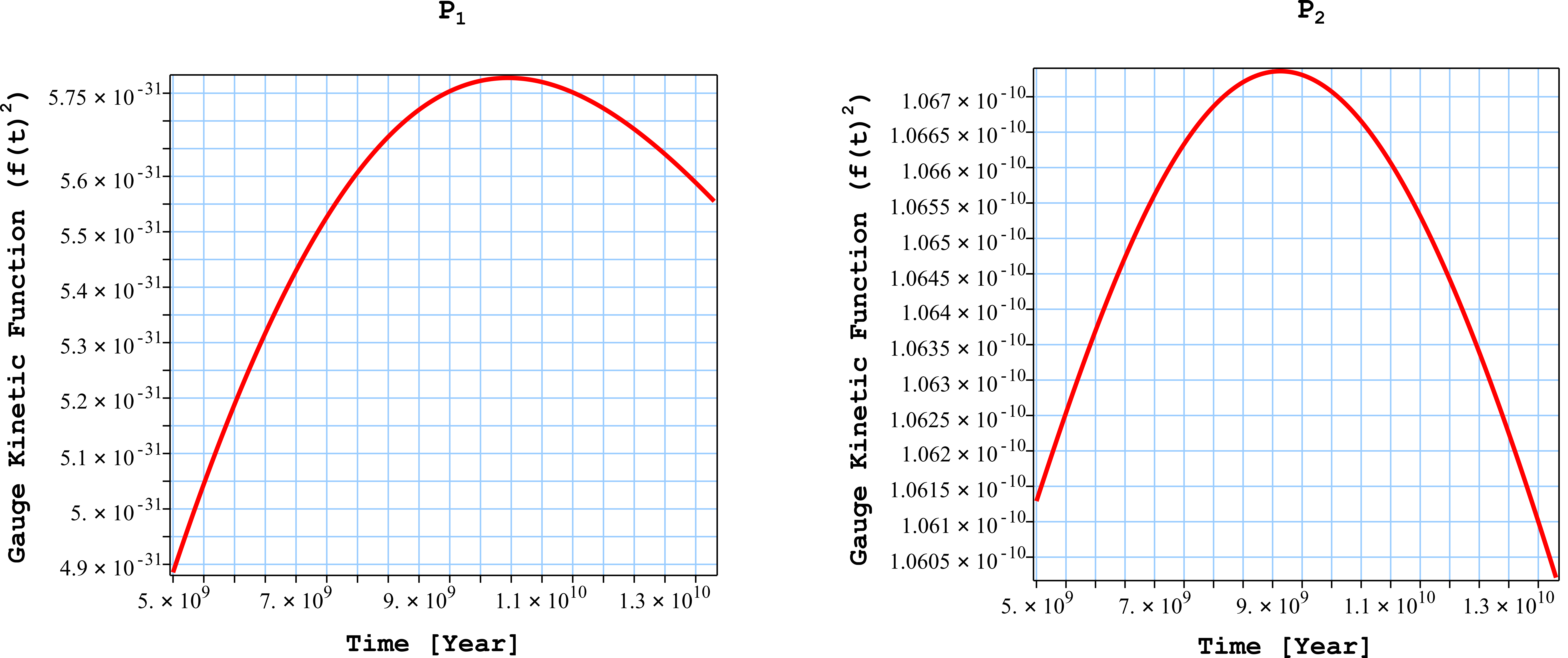}\\
\caption{Plots $\texttt{P}_{1}$ and $\texttt{P}_{2}$ indicate the gauge kinetic function, $f^2$, versus time at the time interval $[5\mathrm{G yr}, 13.801 \mathrm{G yr}]$ for \textbf{C1} and \textbf{C4} classes, respectively.}\label{0f-t}
\end{figure*}
In section (\ref{sect3}), four classes (i.e. \textbf{C1}, \textbf{C2}, \textbf{C3}, and \textbf{C4}) of solutions were obtained. Due to (\ref{AMU01}) and (\ref{AMU02}) or equivalently because of the existences of $n_{j}$ and $k_{j}$ in the solutions, each of these classes of solutions has two sub-classes. Therefore, eight sets of solutions were acquired. In this section, data analysis of these solutions to illustrate the descriptions of late-time-accelerated expansion from the perspective of the studied model, are carried out. But, due to some reasonable reasons which are presented in what follows, we excerpt two sets to perform this interesting work. As is clear, our solutions are more general than ref. \cite{g51}, but since the data analysis done in ref. \cite{g51} was for the set of solutions which have the potential of the form $V \sim V_{0} \exp(- \varphi)$ which is about similar to \textbf{C2}-class (See (\ref{phi123}) and (\ref{pot12})--(\ref{cop12})), hence between \textbf{C1} and \textbf{C2} classes, we analyze \textbf{C1}-class only. Between \textbf{C3} and \textbf{C4} classes, we select \textbf{C4}-class because it seems easier to tune than \textbf{C3}, because we must have a real scale factor and as is observed from (\ref{solC3-C4-2}), there is a power $(1/(m+2)) \approx 1/3$ hence the term under it must be positive, on the other hand, $\sinh^{3/2}(t)$-function grows so quickly than $t^2$-function, hence tuning the constant parameters by choosing $\theta=-1$ (i.e. \textbf{C3}-class) in order to have a real scale factor is more easier than $\theta=+1$ (i.e. \textbf{C4}-class). Note that both are doable, but only for making our work easy we act on this manner. Therefore, up to now, we singled out four sets of solutions among eight sets. Based on the recent observational data, in ref. \cite{beh20192}, it has been demonstrated that $m$ is very close to one\footnote{$\left(\frac{\sqrt{3}-2N_{0}}{\sqrt{3}+N_{0}} \right) \leq m \leq \left(\frac{\sqrt{3}+2N_{0}}{\sqrt{3}-N_{0}} \right)$ in which $N_{0}$ is about $10^{-10}$. Hence, $m$ has a very narrow bound around one.}. The case $m=1$ is FRW. Indeed taking $m=1$ causes that $k_{1}=k_{2}/\sqrt{2}$ which implies only one form for the 4-vector (i.e. $A_{\mu}=(\chi(t);k_{1}A(t),k_{1}A(t),k_{1}A(t))$) instead of (\ref{AMU01}) and (\ref{AMU02}). Note that our solutions for this case holds true and taking each admissible value for $m$ excluding one, makes no considerable changes in the values of parameters and also in plots. Hence, without loss of generality, let us take $m=1$ and consequently $n_{1}=n_{2}=n=1/3$. Indeed, with this choice, four remained cases reduced to two cases. Therefore, our analysis in what follows would be on \textbf{C1} and \textbf{C4} classes by taking $m=1$. Roughly speaking of constant parameters, the forms of the scale factors for \textbf{C1} and \textbf{C2} are totally the same and also the same situation exists between \textbf{C3} and \textbf{C4} classes (See (\ref{Rescalefactor}) and (\ref{solC3-C4-1})). However, other things like the forms of the potentials, coupling functions, scalar fields are different among the obtained cases, but the role of the form of the scale factor is very important than others when we focus on the elaborations of the recent discoveries like late-time-accelerated expansion and phase crossing. Therefore, our selections are completely justifiable and reasonable. Moreover, for these two selected cases, we present thirteen figures ($26$ plots), hence if we analyze all the obtained cases, then $104$ plots should be presented which is completely unusual.\\
First of all, let us review the important values of parameters especially from Planck data 2018 \cite{Planck2018}:
\begin{itemize}
	\item The present value of the scale factor = $1$,
	\item The present value of the redshift = $0$,
	\item The age of the universe = $13.801 \pm 0.024 \; \mathrm{Gyr}$,
	\item The present value of the Hubble parameter = $67.4 \pm 0.5 \; \mathrm{Km.s^{-1}.Mpc^{-1}}$,
	\item The present value of the EoS parameter = $-1.03 \pm 0.03$,
	\item The present value of the temperature of our universe = $2.7255 \pm 0.0006 \; \mathrm{Kelvin}$.
	\item  The onset of acceleration around the redshift $z=0.6$.
\end{itemize}

Using $\mathfrak{B}\text{-function}$ method which has recently been suggested by the author \cite{hashem}, the amounts of the constants parameters are singled out as follows:
\begin{enumerate}
	\item For \textbf{C1}-class ($\lambda=+1$, $c_{2}=0$, $c_{1} \neq 0$):
		\begin{align}
&c_{1}=1/3, \quad c_{4}^2 =1.434699449 \times 10^{-8}, \quad c_{5}=3 \times 10^{-14},  \nonumber \\ &c_{9}=7.103241963 \times 10^{-28}, \quad c_{10}=12623.63242, \nonumber \\ &c_{12}=3.148020216 \times 10^{-11}, \quad c_{13}=-3.134799240 \times 10^{-21}, \quad c_{15}=2.130972589 \times 10^{-41}, \nonumber \\ &c_{16}=-1.044933140 \times 10^{-7}, \quad c_{17}=1049.340072, \quad V_{0}=-9.470989284 \times 10^{-14}, \nonumber \\ & f_{0}=10^{-14}, \quad c_{3}=c_{6}=c_{8}=c_{11}=c_{14}=0.
		\end{align}
	\item For \textbf{C4}-class ($\theta=-1$, $c_{1}=-c_{2}=1/2$, $\delta_{1}=0$, $\delta_{2}=1$):
		\begin{align}
&c_{20}=6.222920245 \times 10^{-11}, \quad c_{21}=0.05150508161, \quad c_{22}=-7.885378233 \times 10^{-11}, \nonumber \\ &c_{23}=0, \quad c_{24}=0.7222580092, \quad c_{25}=9.150720036 \times10^{-11}, \nonumber \\ &c_{26}=0.7080797491, \quad f_{0}=10^{-5}, \quad  V_{0}=6.863040027 \times 10^{-11}.
		\end{align}
\end{enumerate}
With these selections, thirteen figures (Twenty-six plots) are presented for data analysis. In all figures, the left-hand side plots ($\verb"P"_{1}$s) are of \textbf{C1}-class while the right-hand side plots ($\verb"P"_{2}$s) are of \textbf{C4}-class. Both plots in figure (\ref{0a-t}) indicate the scale factors, of increasing characters, expressing first the decelerated and then the accelerated expansion of the universe. According to these, if as usual one sets the present amount of the scale factor to one, $b_{0}=1$, then the age of the universe is found to be $t_{0}=13.801 \; \mathrm{Gyr}$ in both cases. The scale factor versus redshift plots presented in figure (\ref{0a-z}), confirm that the present value of the scale factor and redshift are exactly $1$ and $0$, respectively (i.e. $(b_{0},z_{0})=(1,0)$) for both classes. Also, figure (\ref{0a-z}) indicates that the redshifts go down, while the scale factors increase with time. As usual, ignoring a small variation of the prefactor, we consider that the CMBR temperature falls as $b^{-1}$, then, according to figure (\ref{0T-z}), its present value in both cases would be $T_{0}=2.7255 \; \mathrm{Kelvin}$ (i.e. $(T_{0},z_{0})=(2.7255,0)$). However, as is clear from figure (\ref{0T-z}), getting colder process in $\verb"P"_{2}$-plot is faster than $\verb"P"_{1}$-plot at the same redshift interval. It means that the scale factor of the \textbf{C4}-class grows faster than the scale factor of \textbf{C1}-class at the same time/redshift interval. This point is readily observed from figures (\ref{0a-t}) and (\ref{0a-z}) as well. One may also learn this point by exploiting the Taylor expansion for the forms of the scale factors. The evolutions of the Hubble parameters shown in figure (\ref{0H-t}) demonstrate their behaviors with decreasing natures versus time, as we expect.
The present amount of the Hubble parameter in both cases is $H_{0}=67.40 \; \mathrm{Km.s^{-1}.Mpc^{-1}}$. Figure (\ref{0q-z}) indicates passing from positive to negative values, which corresponds first to the decelerating universe, $q>0$, then to the accelerating universe, $q<0$, and its present value is $q_{0}=-1.045$ for both cases. Obviously, $q=0$ renders the inflection point (i.e. shifting from decelerated to accelerated expansion). According to these plots, the onset of acceleration for \textbf{C1} class is $z_{\mathrm{acc.}}=0.5921$ or equivalently $t_{\mathrm{acc.}}=6.09770 \; \mathrm{Gyr}$ and for the \textbf{C4} one is $z_{\mathrm{acc.}}=0.6183$ which corresponds to $t_{\mathrm{acc.}}=6.96689 \; \mathrm{Gyr}$. Therefore, both estimate that the time of the start of acceleration has been at about half the age of the universe. As is clearly observed from
figure (\ref{0q-z}), the deceleration parameter of \textbf{C1}-class falls quicker than the deceleration parameter of \textbf{C4}-class.
This may be learned from figure (\ref{0H-t}) as well because, as is seen, the speed of the Hubble parameters in $\verb"P"_{2}$-plot of figure (\ref{0H-t}) varies faster than the Hubble parameter of $\verb"P"_{1}$-plot. The low speed of the variation of the Hubble parameter of \textbf{C1}-class causes that its related deceleration parameter survives longer time in the accelerated era than the deceleration parameter of \textbf{C4}-class and it is the reason of the differences between the depths of the holes in graphs before the present time (i.e. The fast speed of the variation of the deceleration parameter, or equivalently, the low speed of the variation of the Hubble parameter $\longrightarrow$ Deep hole and high curvature in graph). The redshift corresponded to the onset of the acceleration of the universe expansion is also recovered by figure (\ref{0H-z}) in which the comoving Hubble parameter as a function of redshift --- i.e. $H(z)/(1+z)$ --- is plotted for both classes; in the base of our model of study, clearly showing the onset of acceleration at $z_{\mathrm{acc.}}=0.5921$ and $z_{\mathrm{acc.}}=0.6183$ for \textbf{C1} and \textbf{C4} classes, respectively. Furthermore, from figure (\ref{0H-z}), it is also learned that the present value of the Hubble parameter is $H_{0}=67.40 \; \mathrm{Km.s^{-1}.Mpc^{-1}}$ in both classes of study. The behaviors of $W_{\mathrm{eff.}}$s in figures (\ref{0W-t}) and (\ref{0W-z}) indicate that the crossing of the phantom divide line $W_{\mathrm{eff.}}=-1$ occurs for both classes from the quintessence phase $W_{\mathrm{eff.}}>-1$ to the phantom phase $W_{\mathrm{eff.}}<-1$. These phase transitions occur at about $(t,z)=(7.128 \mathrm{Gyr},0.526)$ and $(t,z)=(9.578 \mathrm{Gyr},0.332)$ for \textbf{C1} and \textbf{C4} classes, respectively and hence our universe is currently in phantom phase. Therefore, the phase transition of \textbf{C1}-class is befallen sooner than \textbf{C4}-class. Moreover, these transitions occur at redshift-distances $\Delta z=0.066$ and $\Delta z=0.286$ after the onset of acceleration for \textbf{C1} and \textbf{C4}, respectively. The present value of the EoS parameter is calculated to be $W_{\mathrm{eff.}}=-1.03$ for both classes of study. The reason for the deeper holes and higher curvatures in $\verb"P"_{1}$-plots than $\verb"P"_{2}$-plots of figures (\ref{0W-t}) and (\ref{0W-z}) may be argued in the same way as above discussed. When $(\mathrm{d}H / \mathrm{d}t)>0$, then the type of the acceleration of our universe is so-called ``super-acceleration'', since the universe not only accelerates but the Hubble parameter is also increasing. According to figure (\ref{0dH-z}), after the redshifts $z=0.526$ and $z=0.332$, for \textbf{C1} and \textbf{C4} classes, respectively, the type of the expansion of our universe from the point of view of the model of study is super-accelerated expansion. Hence, as we expect, the phase of our universe is phantom in throughout of evolution of the type `super-accelerated-expansion'. Furthermore, the current phase and acceleration of our universe are phantom and super-acceleration. It is worthwhile mentioning that the redshift corresponded to the onset of super-acceleration for each class is exactly equal to the redshift of the interring to phantom phase, as it must be. All the important things which we learned up to now, are also recovered by figure (\ref{0B-function}). The red and green lines in them are the line of inflection points namely shifting from decelerated expansion to accelerated expansion, and Phantom divide line (Phase transition line), respectively. The era $\mathfrak{B}[t,0;a]>0$ represents the expansion of the universe. The region in which $\mathfrak{B}[t,0;H]>0$ is related to super-accelerated expansion. Hence, the blue curves in figure (\ref{0B-function}), demonstrating the evolution of our universe, provide all the important events deduced above through several plots. According to figure (\ref{0p-t}), both scalar fields first increase and then decrease as the universe ages. A difference between the two is that the amounts of the scalar field of \textbf{C1}-class are negative while for the \textbf{C4}-class they are positive. The plots of figure (\ref{0V-t}) indicate the manners of the scalar potentials versus time. As we observe, the behavior of the potential of \textbf{C1}-class first is detractive and then is additive while the potential of the \textbf{C4}-class behaves inverse of \textbf{C1}-class. And finally, the plots of the gauge kinetic functions ($f^2$s) with respect to time are presented in figure (\ref{0f-t}). As is observed their graphs are similar to their own scalar field plots (i.e. first increasing and then decreasing). In a nutshell, without assuming $m=1$, four types of well-known potentials were obtained (Note that $\mu >0$):
\begin{enumerate}
	\item The first case:
	\begin{align}\label{0pott01}
	V(\varphi)=V_{0} \exp(+2 \mu \varphi).
	\end{align}
	\item The second case:
	\begin{align}\label{0pott02}
	V(\varphi)=V_{0} \exp(-2 \mu \varphi).
	\end{align}
	\item The third case:
	\begin{align}\label{0pott03}
	V(\varphi)=V_{0} \cosh^{2}(\mu \varphi).
	\end{align}
	\item The fourth case:
	\begin{align}\label{0pott04}
	V(\varphi)=V_{0} \sinh^{2} (\mu \varphi).
	\end{align}
\end{enumerate}
The first case is the increasing exponential potential with respect to the scalar field while the second case is the decreasing exponential potential with respect to the scalar field. As we know, the unified dark matter potential is of the form $V(\varphi)=V_{0} \left[1+\cosh^{2}(C\varphi) \right]$ or equivalently $V(\varphi)=V_{0} \left[2+\sinh^{2}(C\varphi) \right]$. Hence, roughly speaking of some little differences, it may be argued that the third and fourth cases of the potential are the unified dark matter potentials which are so interesting. These nice results are due to our different taken procedure in sub-section (\ref{sub-section2}) in which we combine two symmetries in two different special ways. We do not present plots of the scalar potentials and the coupling functions versus the scalar field because their behaviors are obvious.\\

\noindent\hrmybox{}{\section{Wheeler-De Witt (WDW) Equation \label{WDW-sect}}}\vspace{5mm}

Let us proceed with the Lagrangians $\mathcal{L}_{1j,2j}$ (\ref{ele1101}) and $\mathcal{L}_{3j,4j}$ (\ref{15sppa}). Consequently, their corresponding Hamiltonians may easily become
\begin{align}
\mathcal{H}_{1j,2j} &= \frac{-1}{l_{3}} \Pi_{u} \Pi_{w}+\frac{n_{j}}{2l_{1}k^{2}_{j}} u^{-2n_{j}-1} \Pi^{2}_{A} - \frac{l_{2}}{2} u^2, \label{wdw1.1}\\
\mathcal{H}_{3j,4j} &= \frac{1}{4 \theta l_{5}} \Pi^{2}_{w}- \frac{1}{4 \theta l_{5}} \Pi^{2}_{u}+\frac{1}{4l_{6j}}u^{-2n_{j}} \Pi^{2}_{A}-l_{4} u^{2}, \label{wdw1.2}
\end{align}
where $\{\Pi_{\kappa}=\partial \mathcal{L}_{*}/\partial \dot{Q}^{\kappa}; \; Q^{\kappa} \in \{w,u,v=A\}\}$  are the conjugated momenta of the configuration space of cyclic coordinates. Note that $\Pi_{\kappa}$s in (\ref{wdw1.1}) are not equal to $\Pi_{\kappa}$s in (\ref{wdw1.2}) because in the former one they are obtained by $\Pi_{\kappa}=\partial \mathcal{L}_{1j,2j} / \partial \dot{Q}^{\kappa}$, while in the latter one they are given by $\Pi_{\kappa}=\partial \mathcal{L}_{3j,4j} / \partial \dot{Q}^{\kappa}$. Furthermore, (\ref{wdw1.1}) and (\ref{wdw1.2}) denote eight number of Hamiltonians, hence, for example, $\Pi_{w}$ in (\ref{wdw1.1}) indicates four number of different $\Pi_{w}$s. \\
By a straightforward canonical quantization procedure, one arrives at
\begin{align}
& \Pi_{\kappa} \to \hat{\Pi}_{\kappa} = - \hat{i} \partial_{\kappa},  \label{s0cqn}\\
& \mathcal{H}_{*} \to
\hat{\mathcal{H}}_{*}(Q^{\kappa},-\hat{i}\partial_{Q^{\kappa}}), \label{s0cqn1}
\end{align}
where $\hat{i}=\sqrt{-1}$. Now, the Wheeler-De Witt (WDW) equations (corresponding to (\ref{wdw1.1}) and (\ref{wdw1.2})) are obtained by the use of the Hamiltonian constraints ($\hat{\mathcal{H}}_{*} \left| \Psi(u,w,v=A) \right\rangle=0$):
\begin{align}
\left[\frac{-1}{l_{3}}\left(- \hat{i} \partial_{u} \right)\left(- \hat{i} \partial_{w} \right)+\frac{n_{j}}{2l_{1}k^{2}_{j}} u^{-2n_{j}-1} \left(- \hat{i} \partial_{A} \right)^{2}-\frac{l_{2}}{2}u^{2} \right]\left| \Psi(u,w,v=A) \right\rangle=0, \label{weel01} \\
\left[\frac{1}{4 \theta l_{5}} \left(- \hat{i} \partial_{w} \right)^{2} - \frac{1}{4 \theta l_{5}} \left(- \hat{i} \partial_{u} \right)^{2}+ \frac{1}{4 l_{6j}}u^{-2n_{j}} \left(- \hat{i} \partial_{A} \right)^{2} - l_{4} u^{2} \right]\left| \Psi(u,w,v=A) \right\rangle=0,   \label{weel02}
\end{align}
where $\left| \Psi(u,w,v=A) \right\rangle$s are the wave functions of the universe. Pursuing the symmetries emerged by the Noether's approach if we use the following conserved currents:
\begin{align}\label{c0c1368}
\Pi_{w}=\Sigma_{1}, \qquad \Pi_{A}=\Sigma_{2},
\end{align}
for all cases of study, then according to \cite{sigma001}
\begin{align}\label{c1c1368}
\left| \Psi \right\rangle = \sum_{\mu=1}^{\nu} \exp \left[ \hat{i} \Sigma_{\mu}Q^{\mu}  \right] \left| \mathcal{F}(Q^{l}) \right\rangle, \quad \nu < l \leq \vartheta,
\end{align}
where $\nu$ is the number of symmetries, $l$ are the directions where symmetries do not exist, $\vartheta$ is the total dimension of the minisuperspace, we obtain a unified form of the wave function for all cases of study:
\begin{align}\label{c2c1368}
\left| \Psi(u,w,v=A) \right\rangle = e^{\hat{i} \Sigma_{1} w} e^{\hat{i} \Sigma_{2} A} \left| \Theta(u) \right\rangle.
\end{align}
It is worthwhile mentioning that the appearance of the exponential functions is due to the separation of variables in eqs. (\ref{weel01}) and (\ref{weel02}) and the quantum versions of the constraints (\ref{c0c1368}), namely
\begin{align}\label{c3c1368}
-\hat{i} \; \partial_{w} \left| \Psi(u,w,A) \right\rangle = \Sigma_{1} \left| \Psi(u,w,A) \right\rangle, \qquad -\hat{i} \; \partial_{A} \left| \Psi(u,w,A) \right\rangle = \Sigma_{2} \left| \Psi(u,w,A) \right\rangle.
\end{align}
Putting the solution (\ref{c2c1368}) in eq. (\ref{weel01}) yields
\begin{align}\label{c4c1368}
\left| \Psi(u,w,v=A) \right\rangle = b_{0} \; e^{\hat{i} \Sigma_{1} w} \; e^{\hat{i} \Sigma_{2} A} \; e^{b_{1} u^{3}+b_{2}u^{-2n_{j}}},
\end{align}
where $b_{1}=- \hat{i} l_{2} l_{3} / 6$, $b_{2}=- \hat{i} l_{3} /(4 l_{1} k^{2}_{j})$, and $b_{0}$ is an integration constant. As is clearly observed, the oscillating feature of the wave function of the universe recovers the so-called Hartle criterion \cite{HARTEL1}.\\
Unfortunately, inserting the solution (\ref{c2c1368}) into eq. (\ref{weel02}) does not give an analytical solution. Nonetheless, by the numerical methods it may easily be demonstrated that it also leads to solutions that have oscillating feature, hence the Hartle criterion is also recovered.\\

\noindent\hrmybox{}{\section{Conclusion}}\vspace{5mm}

In this paper, a modified teleparallel gravity action, containing gauge fields as a substantial part, in the framework of teleparallel gravity, was investigated. The background geometry of the studied model was anisotropic and homogeneous which covers FRW as well. By the use of the Noether approach, eight sets of solutions, leading to late-time-accelerated expansion and phase crossing from quintessence to phantom, were acquired.\\
A usual way in dealing with total symmetry generator produced by Noether symmetry approach is that we split it into sub-symmetries and then select some of them, yielding suitable analytical solutions, to be carried by the solutions. But this approach puts limitation on the forms of the unknown functions obtained by Noether symmetry approach. In order to have more suitable solutions and interesting forms for unknown functions, an approach, CSSS-trick, was suggested. Using this trick, in our case of study, the unified dark matter potentials of the forms $V=V_{0} \cosh^{2}(\mu \varphi)$ and also $V=V_{0} \sinh^{2}(\mu \varphi)$ were produced which are highly rewarding.\\
Utilizing the $\mathfrak{B}\text{-function method}$, data analysis of the results was performed. It was demonstrated that our solutions completely conform with all important events and astrophysical and observational data and consequently, the resulting cosmological model accommodate all the important events and data. For example, \emph{some} of our findings and astrophysical data are compared in Table (\ref{TTT}).\\
\begin{table*}
\caption{Comparison of \emph{some} of our findings (\textbf{C1} and \textbf{C4} classes) and observational data.} \label{TTT}
\centering
\begin{tabular}{p{2.5 in} p{1.6 in} p{1.2 in} p{1.2 in}}
\toprule[1.7pt]
\textbf{Parameter} &\textbf{Astrophysical data 2018} & \textbf{Our findings (C1)} &\textbf{Our findings (C4)}
\\
\toprule[1.4pt]
The present value of the scale factor & $1$ & $1$ & $1$  \\
The present value of the redshift & $0$ & $0$ & $0$ \\
The age of the universe & $13.801 \pm 0.024 \; \mathrm{Gyr}$ & $13.801 \; \mathrm{Gyr}$ & $13.801 \; \mathrm{Gyr}$ \\
The present value of the Hubble parameter & $67.4 \pm 0.5 \; \mathrm{Km.s^{-1}.Mpc^{-1}}$ & $67.4 \; \mathrm{Km.s^{-1}.Mpc^{-1}}$ & $67.4 \; \mathrm{Km.s^{-1}.Mpc^{-1}}$ \\
The present value of the EoS parameter & $-1.03 \pm 0.03$ & $-1.03$ & $-1.03$ \\
The present value of the temperature & $2.7255 \pm 0.0006 \; \mathrm{Kelvin}$ & $2.7255 \; \mathrm{Kelvin}$ & $2.7255 \; \mathrm{Kelvin}$ \\
The redshift of the onset of acceleration & around $0.6$ & $0.5921$ & $0.6183$\\
\toprule[1.7pt]
\end{tabular}
\end{table*}
Finally, using the Wheeler-De Witt (WDW) equation, it was indicated that the Hartle criterion due to the oscillating feature of the wave function of the universe is recovered.

Pursuant to some papers, the studied model is successful in the description of the early inflation, hence we conclude that gauge fields are able to produce both early inflation and late-time-accelerated expansion and consequently, through this term, a unified model describing all stages of the universe may be achieved.\\

\section*{\noindent\goldmybox{red}{\vspace{3mm} Acknowledgments \vspace{3mm}}}

This work has been supported financially by Research Institute for Astronomy $\&$ Astrophysics of Maragha (RIAAM) under research project No. 1/5440-33.\\

\hrule \hrule \hrule \hrule \hrule \hrule

\end{document}